\documentclass[reprint, superscriptaddress,
amsmath,amssymb,
aps,
prb,
floatfix,
]{revtex4-2}

\usepackage{graphicx}
\usepackage{float} 
\usepackage{bm}
\usepackage{array}%
\usepackage{comment}
\usepackage{color}
\usepackage{multirow}

\definecolor{ForestG}{rgb}{0.31, 0.78, 0.47}
\definecolor{MyGreen}{RGB}{54,165,54} 
\definecolor{violet}{RGB}{127, 0, 255}

\begin{document}

\preprint{APS/123-QED}

\title{Developing a valence force field model for wurtzite semiconductors by exploiting similarities with [111]-oriented zinc blende systems: The case of wurtzite boron nitride, III-N materials and (B,In,Ga)N alloys}

\author{Aisling Power}
\email{aisling.power@tyndall.ie}
\affiliation{Tyndall National Institute, Lee Maltings, Dyke Parade, Cork T12 R5CP, Ireland}
\affiliation{School of Physics, University College Cork, Cork T12 YN60, Ireland}
\author{Cara-Lena Nies}
\affiliation{Tyndall National Institute, Lee Maltings, Dyke Parade, Cork T12 R5CP, Ireland}
\author{Stefan Schulz}
\affiliation{Tyndall National Institute, Lee Maltings, Dyke Parade, Cork T12 R5CP, Ireland}
\affiliation{School of Physics, University College Cork, Cork T12 YN60, Ireland}

\date{\today}

\begin{abstract}
Controlling the crystal phase and lattice mismatch of semiconductors offers a powerful route to engineer electronic and optical properties of heterostructures. As a consequence, semiconductors in the wurtzite phase are increasingly sought after, superseding the thermodynamically favored cubic zinc blende phase. Empirical atomistic modeling, required for large scale simulations of heterostructures and their properties, relies heavily on valence force field (VFF) methods to find the equilibrium atomic positions in an alloy. For zinc blende crystals, VFF models are well-established. In the case of wurtzite, such parameters are frequently adopted without rigorous analysis, despite subtle but consequential differences from the zinc blende structure. Such an approach can compromise accuracy in describing material properties, since the structural differences between zinc blende and wurtzite directly influence electronic and optical characteristics. Based on the analytical VFF model by Tanner \emph{et al.}~\cite{TaCa2019}, and using structural similarities between wurtzite and [111]-oriented zinc blende, we construct a wurtzite VFF without introducing additional parameters. Our framework relies on analytic expressions and minimization routines to project zinc blende models onto wurtzite systems. Beyond elastic tensors, we train the model to reproduce bond length asymmetries and band gaps by using output of the VFF model in density functional theory calculations. Applied to wurtzite III-N compounds and BN, the model accurately reproduces targeted observables but also properties it has not been trained on, including the internal parameter $u$. We further validate the model on highly mismatched alloys such as (B,Ga)N and (B,In,Ga)N, exhibiting good agreement between VFF and density functional theory results when using identical supercells in these calculations.  

\end{abstract}

\maketitle

\section{\label{sec:intro}Introduction}

Semiconductor heterostructures, such as quantum wells (QWs) and quantum dots (QDs), are attracting significant research interest, as they can facilitate realization of fundamental 'textbook' quantum physics~\cite{Simo2017, BaCa1999, FoIs2007}. In addition to these fundamental physics aspects, semiconductor heterostructures have paved the way for many applications, from conventional optoelectronic devices like light-emitting diodes to non-classical light sources such as single-photon emitters~\cite{ODonnell_SemiconductorMaterials, Upadhyay_AlGaN_HEMTSensors, Low_Dim_QuantumSys, Tidrow_QWIPs, UVRoadMap2020}.

To tailor the electronic and optical properties of QWs or QDs for (device) applications, material composition, well width or dot size and shape are predominantly targeted \cite{MiSc2022,KuRa2025, ScCa2015}. However, fundamental material properties, for instance, the band gap value and its nature (indirect or direct), may limit the degree to which a material underlying a heterostructure is useful for device applications. As a consequence, research efforts have focused on the development and prediction of new material, alloy and heterostructure systems~\cite{WaZi2020, MaDm2017, ChLi2020, LoJi2018}. Here, research attention has shifted toward controlling the crystal phase of well-established semiconductors to enable novel devices, including light emitters based on Group-IV systems compatible with complementary metal-oxide semiconductor technologies~\cite{SuRo2021}.

Recent prominent examples include wurtzite (lonsdaleite) Ge~\cite{SuRo2021} and SiGe~\cite{DePr2014, PevLa2024}, or in terms of III-V materials, wurtzite (WZ) (Al,Ga)As semiconductors~\cite{LeRe2020,SiPr2023}. For example, cubic zinc blende (ZB) AlAs~\cite{Adac1994, JiXi2018} is an indirect band gap material and as such has limited potential as the active region in optoelectronic devices. The same is true for Ge or Si, which are both indirect semiconductors in the diamond phase.  However, theoretical and experimental studies have highlighted that WZ (Al,Ga)As ~\cite{DePr2010, LeRe2020} or lonsdaleite Ge~\cite{ZhIq2000, BhAk2024, DePr2014,LeRe2020,SiPr2023}  exhibit a direct band gap, which could be used in, e.g., lasers and optical amplifiers for integrated photonics~\cite{Sore2006}.

As well as band gap tailoring, controlling the crystal phase can be used to engineer electric polarization or strain fields in nanostructures. The latter aspect 
is important for reducing defect densities and thus non-radiative recombination centers in QWs. A recent example of managing strain and polarization fields is the 
use of WZ boron nitride (BN) in WZ III-N QWs, e.g., adding BN to gallium nitride (GaN) or aluminium nitride (AlN) systems~\cite{KuRo2020, GaOr2010, PaAh2016, 
PaAh2019, LoKi2019}, given the smaller lattice constant of WZ BN when compared to AlN, GaN or indium nitride (InN). However, in addition to the fact that BN is 
most stable in the layered hexagonal and cubic phases~\cite{SoTu1999} and tends to crystallize preferentially in the cubic phase~\cite{GuMo2017} when grown 
alongside WZ III-N materials, boron containing III-N alloys can exhibit extremely large local lattice mismatches. Taking WZ BN and WZ GaN as an example, the 
lattice mismatch between these two materials is approximately 20\%. Experimental studies have shown that incorporating WZ BN into III‑Nitrides is typically 
limited to 15–20\% BN content before significant crystal structure degradation occurs. High‑quality single‑phase films are generally achievable only up to $\approx$10\%
BN, beyond which phase separation, defect formation, and poor surface morphology become dominant~\cite{GaOr2010,LiLi2020, TrLi2020,SaMs2021}.

Beyond growth challenges, highly mismatched alloys (HMAs) display structural, electronic, and optical properties that deviate markedly from conventional systems such as (Al,Ga)As~\cite{BrOR2024, JaSp2019, YuNo2014, ChMi2019}. This necessitates the use of atomistic electronic-structure simulations for accurately predict heterostructure behavior~\cite{BrOR2024}. While density functional theory (DFT) provides a first-principles framework, capturing electronic states localized in the band gap, alloy disorder in QWs, or experimentally relevant dilute compositions ($<$5\%) often demand supercells beyond the reach of conventional DFT. To overcome this, empirical atomistic methods, benchmarked against smaller-scale DFT calculations, enable efficient large scale simulations ($\geq$ 50,000 atoms) at reduced computational cost~\cite{BrOR2024}.

A central element of empirical atomistic models is the valence force field (VFF), which determines equilibrium atomic positions in alloys and heterostructures. VFFs rely on force constants describing interatomic interactions such as bond stretching and bending, typically obtained by fitting to the elastic tensor of a material which reflects its crystal symmetry. Though, transferability across crystal structures, e.g., from ZB to WZ, is not guaranteed. For ZB semiconductors, analytical relations between force constants, elastic constants, and the internal strain parameter are available~\cite{TaCa2019}. In contrast, WZ systems are considerably more complex due to their larger unit cell and the greater number of independent elastic constants and strain parameters~\cite{Ny1985,Caro2012}. For crystal‑phase heterostructures, or WZ inclusions in ZB QWs, such as basal‑plane stacking faults~\cite{BiDa2022}, it is therefore essential to employ a VFF model that treats both phases consistently within a single framework, without introducing extra terms in the VFF potential or parameters. In the absence of such consistency, predictive modeling of polytypic structures remains unreliable, limiting both theoretical insight and practical device design.

Thus, in this work, we construct a WZ VFF potential by extending the ZB model of Ref.~\cite{TaCa2019}. The Tanner \emph{et al.}~\cite{TaCa2019} approach is advantageous because (i) all parameters are derived directly from elastic constants, without numerical fitting, and (ii) it applies to both polar and non‑polar materials. Exploiting the analytic ZB expressions and the close correspondence between the elastic tensors of WZ and [111]‑oriented ZB, we develop the WZ VFF without the introduction of additional parameters. Furthermore, unlike prior studies on WZ systems~\cite{GrNe2001,CaNi2010}, we fit not only to the elastic constants and bond asymmetry but also band gaps, using VFF‑optimized atomic positions within DFT electronic structure calculations. Our procedure is fully general, applicable to both polar and non‑polar systems, and extends the role of VFF models beyond mechanical consistency to direct electronic accuracy. Our method advances VFF model development beyond mere elastic fitting, achieving electronic predictivity and validating its power at a level usually not considered in conventional approaches. 

Moreover, we apply the VFF model beyond binary compounds, targeting both structural properties (bond‑length distributions and lattice constants) and electronic characteristics (band gaps) of HMAs. The predictive power of the model is validated by direct comparison of supercell calculations with DFT results. This procedure goes well beyond conventional approaches, where VFF models are typically benchmarked only for binary systems and without validation of either bond‑length distributions or electronic structure against DFT.

Specifically, we consider WZ BN together with the ``conventional'' III‑N materials AlN, GaN, and InN. 
For the binary materials, this approach delivers accuracy exceeding 92\% relative to DFT: elastic constants, bond lengths, and band gaps are reproduced within 7\%, while, remarkably, the lattice parameters, internal $u$ parameter, and $c_0/a_0$ ratio are captured to better than 98\% agreement without explicit fitting. 

We extend our VFF model to large alloy supercells of the HMAs (B,Ga)N and (B,In,Ga)N, investigating relaxed atomic positions that explicitly exhibit alloy disorder at the atomistic level. The resulting bond‑length distributions are compared with results from DFT studies performed on the same supercells. Alike the binary systems, the VFF‑relaxed positions, without further optimization, are used directly in DFT band‑gap calculations. Across a variety of alloy compositions and configurations, our VFF approach reproduces lattice parameters with $>$99\% accuracy and band gaps with $>$92\% accuracy. Even in the most challenging cases, e.g. clustering of boron atoms in (B,Ga)N alloys, maximum deviations remain at 7\%, with band gaps within 1\% of full DFT results for random alloys. All this demonstrates the predictive fidelity for HMAs well beyond conventional VFF benchmarks. Moreover, the VFF model can now serve as a robust and reliable foundation for large‑scale electronic structure calculations, thus enabling predictive studies of complex alloys and heterostructures.

The manuscript is organized as follows. Section~\ref{sec:theoretical_framework} introduces the theoretical framework of the VFF model, including (i) the potential employed, Sec.~\ref{subsec:potential}, (ii) a comparison of WZ, [001]- and [111]‑oriented ZB elastic tensors, Sec.~\ref{subsec:elastic_tensor}, and (iii) details of the parameter extraction procedure, Sec.~\ref{subsec:fitting}. Section~\ref{sec:boron} presents results of the VFF model for binary III‑N systems and boron‑containing III-N alloys. We discuss the VFF potential for WZ BN, AlN, InN, and GaN, using DFT calculations, Sec.~\ref{sec:DFT}, for parameterization and validation, in Sec.~\ref{subsec:BN_fitting}. The model is then applied to the HMAs (B,Ga)N, Sec.~\ref{subsec:BGaN}, and (B,In,Ga)N, Sec.~\ref{subsec:BInGaN}, with results compared directly to DFT supercell calculations. Finally, our conclusions are given in Sec.~\ref{sec:conclusion}.

\begin{figure}[t!]
    \centering
    \includegraphics[width=1\linewidth]{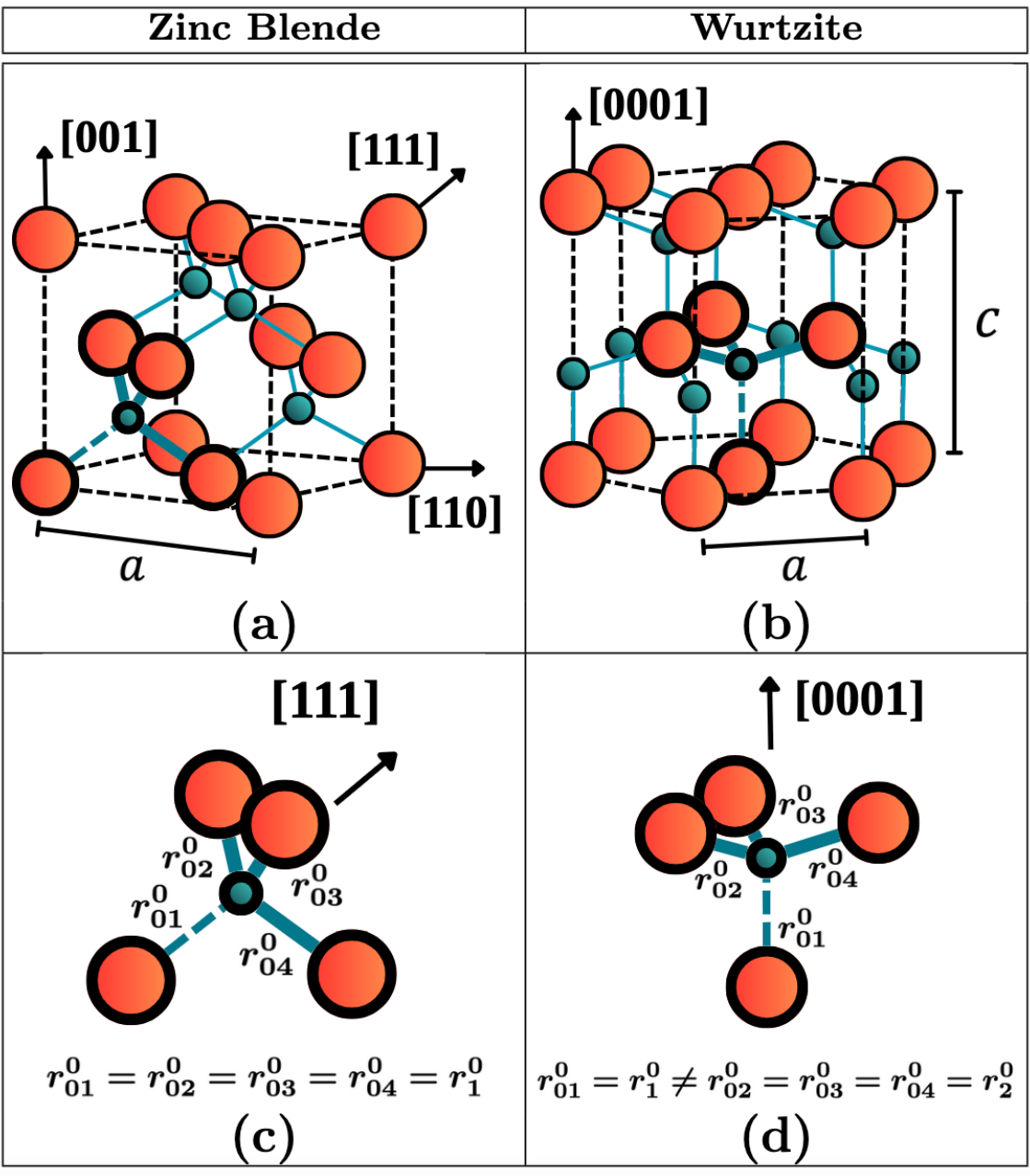} 
    \caption{(a) Zinc blende and (b) wurtzite crystal structures, with crystal axis directions in Miller indice notation. (c) Zinc blende and (d) wurtzite nearest neighbor tetrahedrons, respectively. The equilibrium nearest neighbor bond lengths are denoted by $r^{0}_{01}$ (dashed line), $r^0_{02}$, $r^0_{03}$ and $r^0_{04}$ (solid lines) following notation in the main text.}
    \label{fig:structures}
\end{figure}

\section{Theoretical Framework}
\label{sec:theoretical_framework}

In this section, we present our VFF model development procedure. In Sec.~\ref{subsec:potential}, we detail the key components of the VFF potential, originally introduced by Musgrave and Pople~\cite{MuPo1962}, that underpin our study. To be able to guide the VFF parameterization, we review in Sec.~\ref{subsec:elastic_tensor} the similarities between the elastic tensor of WZ and [111]-oriented ZB, and how to transform the elastic tensor from [001] to [111]-oriented ZB. Finally, details of the VFF parameter extraction are presented in Sec.~\ref{subsec:fitting}. 

\subsection{Interatomic Potential}  
\label{subsec:potential}

The VFF potential first introduced by Musgrave and Pople~\cite{MuPo1962} was subsequently extended by Martin~\cite{Mart1970}. Tanner \emph{et al.}~\cite{TaCa2019} then further employed this potential to formulate a model for polar and non-polar ZB materials, with the force constants derived directly from elastic constants and internal strain parameters, without numerical fitting. We extend this model to the WZ lattice, without introducing additional force constants, with the aim of establishing a unified framework applicable to both ZB and WZ structures. 
To achieve this, it is important to understand the similarities and differences of the WZ and ZB crystal structures. Figure~\ref{fig:structures}
depicts schematic illustrations of ZB and WZ conventional unit cells and nearest-neighbor arrangements, respectively. Based on the tetrahedral bonding with all four nearest neighbor bond lengths being identical in ZB, Fig.~\ref{fig:structures} (c), the potential of each atom $i$ in a ZB crystal reads:    
\begin{widetext}
    \begin{eqnarray}
    \label{eqn:MartinPot}
    \nonumber
    V_i &=& \frac{1}{2} \sum_{j \neq i} \frac{1}{2} k_{r} (r_{ij} - r_{ij}^0)^2\\
    &+& \sum_{j \neq i} \sum_{\substack{k \neq i\\ k > j}} \left[\frac{k_{\theta}^i}{2} r_{ij}^0 r_{ik}^0 (\theta_{ijk} - \theta^0_{ijk})^2 + k_{r \theta}^i [r^0_{ij} (r_{ij} - r^0_{ij}) + r^0_{ik}(r_{ik} - r_{ik}^{0})] (\theta_{ijk} - \theta_{ijk}^0) + k_{rr}^i (r_{ij} - r_{ij}^0) (r_{ik} - r_{ik}^0) \right]\\ \nonumber
    &+&\underbrace{\frac{1}{2} \sum_{j \neq i}' S_{ij}\frac{ e^2}{4 \pi \epsilon_0 r_{ij}}}_{V_\text{att}} - \underbrace{\frac{1}{2} \sum_{j \neq i}^{nn} \frac{1}{4} \alpha_\text{M} S_{ij} \frac{e^2}{4 \pi \epsilon_0 {r_{ij}^{0}}^{2}} (r_{ij} - r_{ij}^0)}_{V_\text{rep}} \,\, .
    \end{eqnarray}
    \end{widetext}
In the above notation, $i$ denotes the central atom in a tetrahedron of nearest-neighbor atoms, here labeled $j$ and $k$, respectively. All sums run over nearest neighbours only, except the $\sum^{'}_{j \neq i}$ which extends over the entire crystal. The length of the bond between atom $i$ and $j$ is denoted by $r_{ij}$, while $\theta_{ijk}$ is the angle between the bonds $r_{ij}$ and $r_{ik}$. The equilibrium bond lengths and angles are labeled as $r_{ij}^{0}$ and $\theta_{ijk}^{0}$, respectively. As already mentioned above, due to the symmetry of the ZB crystal, the lengths of the four nearest neighbor bonds, $r^0_{ij}$, are identical. This means that when denoting the central atom of the tetrahedron formed by the nearest neighbors as $i=0$ one is left with $r^0_{01}=r^0_{02}=r^0_{03}=r^0_{04}$.

The force constants describe the different covalent interactions between the atoms in the lattice. The terms involving $k_r$ characterize the resistance of changing the bond length away from its equilibrium value. Similarly, $k_{\theta}$ captures the resistance to changing the bond angle. The force constant $k_{r\theta}$ represents the combined effect of changes in bond angle and lengths, while $k_{rr}$ captures interactions of neighboring bonds sharing an atom, e.g., if one bond is stretched the other is shortened. The two final terms in Eq.~(\ref{eqn:MartinPot}) describe Coulomb effects, with $S_{ij}=(Z_i^*Z^*_{j})/\epsilon_r$ being a dimensionless effective charge parameter that involves the effective ion charge, $Z^*$,  and the dielectric constant, $\epsilon_r$, of the material; $\alpha_\text{M}$ is the Madelung constant. The first contribution, $V_\text{att}$, is the screened Coulomb attraction, originally introduced by Blackman~\cite{Blac1958}. The second contribution, $V_\text{rep}$, a linear repulsion term, ensures the stability of the crystal at equilibrium and preserves the symmetry of the elastic tensor, further outlined by Martin~\cite{Mart1970}. Following Martin~\cite{Mart1970}, we adopt the simplifying notation (SI units) $S_{ij} = (Z^*)^2/\epsilon_r=S$.

Tanner \emph{et al.}~\cite{TaCa2019} expanded Eq.~(\ref{eqn:MartinPot}) in terms of strain and sub-lattice displacement. In doing so, and connecting this to the elastic energy of the primitive cell, analytic expressions for the force constants and $S$ parameter can be obtained in terms of (i) elastic constants, (ii) internal strain (Kleinmann) parameter, and (iii) the inner elastic constant for a ZB crystal. The corresponding equations are summarized in Appendix~\ref{sec:appendix}. For a WZ crystal structure the procedure becomes far more complex due to the increased number of (i) atoms in the unit cell (ZB: 2 atoms; WZ: 4 atoms), (ii) independent elastic constants (ZB: 3 constants; WZ: 5 constants), and (iii) internal strain parameters (ZB: 1 parameter; WZ: 5 parameters)~\cite{Caro2012}. Furthermore, while in ZB all four bond lengths are the same ($r^0_{01} =r^0_{02}=r^0_{03}=r^0_{04}=r^0_1$), in WZ only three of four are identical, as schematically shown in Fig.~\ref{fig:structures} (d): $r^0_{01}=r^0_1 \neq r^0_{02}=r^0_{03}=r^0_{04}=r^0_2$. The WZ crystal structure therefore exhibits a bond length asymmetry, which must be accounted for when developing and employing a VFF model for WZ on the basis of Eq.~(\ref{eqn:MartinPot}). 

Without such modifications, one can adopt the original ZB version of the potential to achieve a reasonable approximation of a WZ system by exploiting similarities in the crystal structures of [111]-oriented ZB and WZ. For instance, as shown in Ref.~\cite{ScCa2011}, the elastic tensor of a [111]-oriented ZB structure is similar to that of the WZ crystal. 
This similarity, and the analytic expressions for force constants in ZB materials, will guide the development of our WZ VFF model that consistently builds on Eq.~(\ref{eqn:MartinPot}). Before presenting the model, we briefly review the similarities and differences in the elastic tensor of ZB and WZ materials. Moreover, we discuss analytic expressions that transform elastic constants from [001]- to [111]-oriented ZB systems and vice versa, as these provide a starting point for determining VFF force constants that describe the WZ elastic tensor.

\subsection{Elastic Tensor Symmetries}
\label{subsec:elastic_tensor} 

The cubic symmetry of a ZB crystal, when considering an orientation along the [001]-direction, results in an elastic tensor with three independent constants, namely $C_{11}$, $C_{12}$, and $C_{44}$, which reads~\cite{Ny1985}: 

\begin{equation}
    C_\text{ZB}^{(001)} = 
    \begin{pmatrix}
    C_{11}^\text{ZB} & C_{12}^\text{ZB} & C_{12}^\text{ZB} & 0 & 0 & 0\\
    C_{12}^\text{ZB} & C_{11}^\text{ZB} & C_{12}^\text{ZB} & 0 & 0 & 0\\
    C_{12}^\text{ZB} & C_{12}^\text{ZB} & C_{11}^\text{ZB} & 0 & 0 & 0\\
    0 & 0 & 0 & C_{44}^\text{ZB} & 0 & 0 \\
    0 & 0 & 0 & 0 & C_{44}^\text{ZB} & 0 \\
    0 & 0 & 0  & 0 & 0 & C_{44}^\text{ZB}
    \label{matrix:001ZB}
    \end{pmatrix}\, .
\end{equation}

\begin{figure*}[t!]
    \includegraphics[width=1\linewidth]{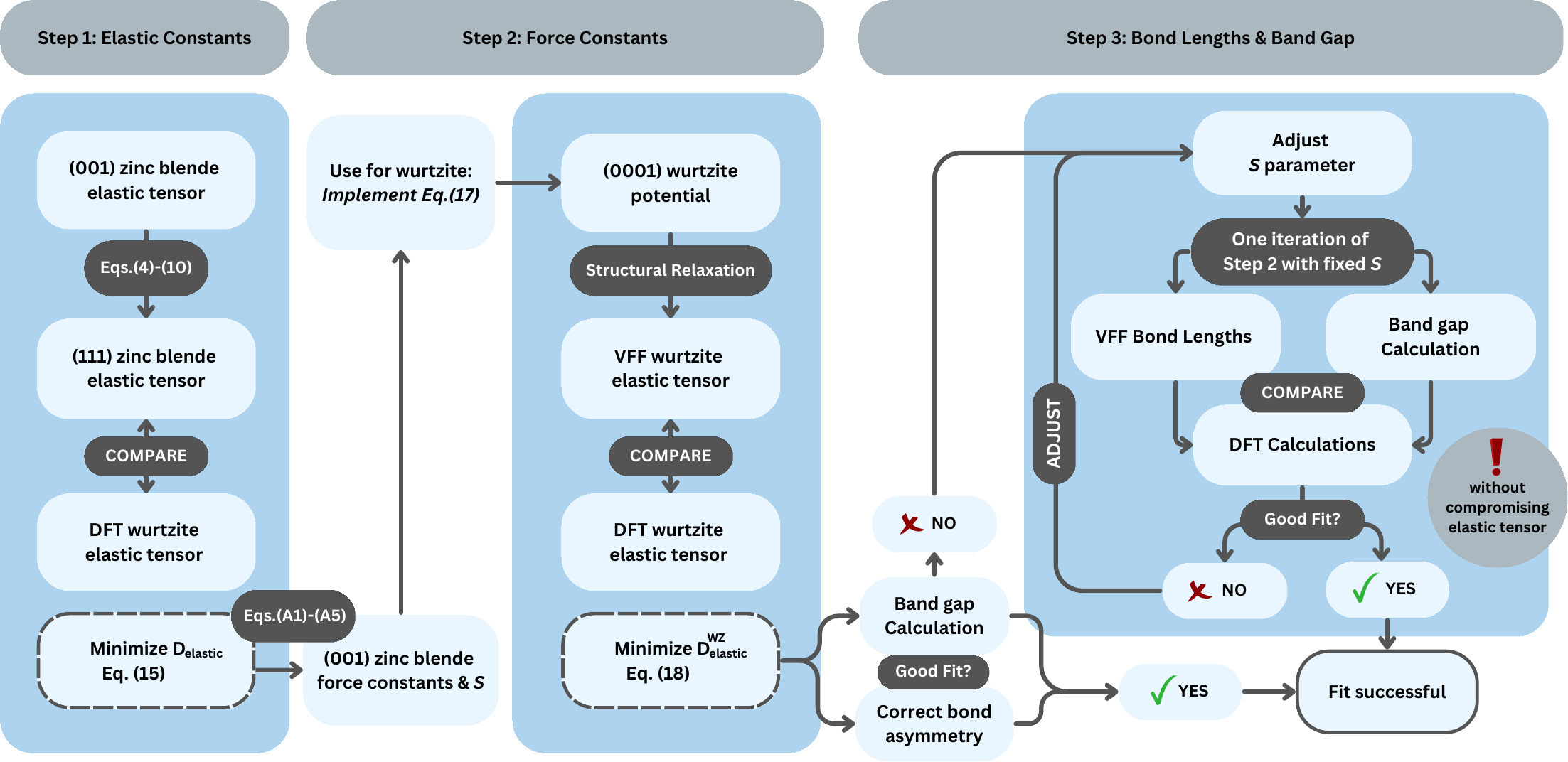}
    \caption{Schematic illustration of the workflow to establish a valence force field (VFF) model for wurtzite semiconductor materials building on the analytic model by Tanner \emph{et al.}~\cite{TaCa2019} for zinc blende materials. More details on the different steps are given in the main text. The last step is introduced specifically for materials where the band gap is a key property for, e.g., device applications.} 
    \label{fig:workflow}
\end{figure*}

As shown in Ref.~\cite{ScCa2011}, when rotating the [001]-elastic tensor to the [111]-direction, it becomes:
\begin{equation}
    C_\text{ZB}^{(111)} = 
    \begin{pmatrix}
    C_{11}^{'} & C_{12}^{'} & C_{13}^{'} & 0 & C_{15}^{'} & 0\\
    C_{12}^{'} & C_{11}^{'} & C_{13}^{'} & 0 & -C_{15}^{'} & 0\\
    C_{13}^{'} & C_{12}^{'} & C_{33}^{'} & 0 & 0 & 0\\
    0 & 0 & 0 & C_{44}^{'} & 0 & -C_{15}^{'} \\
    C_{15}^{'} & -C_{15}^{'} & 0 & 0 & C_{44}^{'} & 0 \\
    0 & 0 & 0  & -C_{15}^{'} & 0 & C_{66}^{'}
    \label{matrix:111ZB}
    \end{pmatrix}\, .
\end{equation}
The elastic constants in the [111]-oriented system can be expressed in terms of $C_{11}$, $C_{12}$ and $C_{44}$ as follows:
\begin{align}
    \label{eqn:001ZBto111ZB_start}
    C_{11}^{'} &= \frac{1}{2} (C_{11}^\text{ZB} + C_{12}^\text{ZB}) + C_{44}^\text{ZB}\, , \\
    C_{33}^{'} &= \frac{3}{2} C_{11}^{'} - \frac{1}{2} C_{12}^{'} - C_{44}^{'}\, , \\
    C_{44}^{'} &= \frac{1}{3} (C_{11}^\text{ZB} + C_{12}^\text{ZB}) + \frac{1}{3} C_{44}^\text{ZB}\, , \\
    C_{12}^{'} &= \frac{1}{6} (C_{11}^\text{ZB} + 5 C_{12}^\text{ZB}) - \frac{1}{3} C_{44}^\text{ZB}\, , \\
    C_{13}^{'} &= - \frac{1}{2} C_{11}^{'} + \frac{3}{2} C_{12}^{'} + C_{44}^{'}\, , \\
    C_{15}^{'} &= \frac{1}{\sqrt{2}} C_{11}^{'} - \frac{1}{\sqrt{2}} C_{12}^{'} - \sqrt{2} C_{44}^{'}\, , \\
    C_{66}^{'} &= \frac{1}{2} (C_{11}^{'} + C_{12}^{'})\, .
    \label{eqn:001ZBto111ZB_end}
\end{align}
Here, $C_{ij}^\text{ZB}$ indicate the [001]-oriented ZB elastic constants, while $C_{ij}^{'}$ denote elastic constants in the [111]-oriented ZB case. If the elastic constants $C_{ij}^{'}$ are known, one can also derive a set of equations that can be used to determine elastic constants $C_{11}^\text{ZB}$, $C_{12}^\text{ZB}$ and $C_{44}^\text{ZB}$: 
\begin{align}
\label{eqn:111ZBto001ZB_start}
    C_{11}^\text{ZB} &= - \frac{3}{2} \left(\frac{1}{3} C_{11}^{'} + C_{12}^{'} - 2(C_{44}^{'} + C_{12}^{'})\right)\, , \\
    C_{12}^\text{ZB} &= \frac{3}{2} \left(\frac{1}{3} C_{11}^{'} + C_{12}^{'} - \frac{2}{3}(C_{44}^{'} + C_{12}^{'})\right)\, , \\
    C_{44}^\text{ZB} &= C_{11}^{'} - \frac{1}{2} \left(C_{12}^\text{ZB} + C_{11}^\text{ZB}\right)\, .
    \label{eqn:111ZBto001ZB_end}
\end{align}
Overall, this means that if one set of elastic constants is given, either $C_{ij}^{'}$ or $C_{ij}^\text{ZB}$, the expression in Appendix~\ref{sec:appendix} can be used to directly calculate the force constants for the VFF potential, Eq.~(\ref{eqn:MartinPot}), that describe the elastic tensor of both [001]- and [111]-oriented ZB systems.   

Moving now to WZ, the elastic tensor is very similar to the [111]-oriented ZB tensor~\cite{Ny1985}:
\begin{equation}
    C_\text{WZ}^{(0001)} = 
    \begin{pmatrix}
    C_{11} & C_{12} & C_{13} & 0 & 0 & 0\\
    C_{12} & C_{11} & C_{13} & 0 & 0 & 0\\
    C_{13} & C_{12} & C_{33} & 0 & 0 & 0\\
    0 & 0 & 0 & C_{44} & 0 & 0 \\
    0 & 0 & 0 & 0 & C_{44} & 0 \\
    0 & 0 & 0  & 0 & 0 & C_{66}
    \label{matrix:WZ}
    \end{pmatrix}\,\, .
\end{equation}
This similarity stems from the nearly identical nearest-neighbor environments in [111]-ZB and WZ, with the main difference at this level being the bond-length asymmetry in WZ discussed in Sec.~\ref{subsec:potential}. Beyond nearest neighbors, of the 12 second-nearest neighbors, 9 share the same arrangement in WZ and [111]-oriented ZB, while the remaining 3 are rotated by $\pi/3$ in ZB relative to WZ~\cite{ScCa2011}. This difference in the second nearest neighbor environment leads to an $ABCABCABC\ldots$ stacking sequence along the $[111]$ direction in ZB, while in WZ the sequence is $ABABAB\ldots$ along the $[0001]$-crystal axis.

Employing a [111]-oriented ZB elastic tensor, closely matching the WZ tensor in both magnitude and ratios of elastic constants, we obtain a set of force constants that reasonably describe WZ materials. Using the expressions for the elastic constants, Eqs.~(\ref{eqn:001ZBto111ZB_start})–(\ref{eqn:111ZBto001ZB_end}), together with Appendix~\ref{sec:appendix}, the force constants and $S$ are determined entirely analytically. Subsequent refinement of the force constants and the potential, Eq.~(\ref{eqn:MartinPot}), is then required to capture the full WZ elastic tensor and bond-length asymmetry. Crucially, this procedure provides an initial set of force constants which yield an elastic tensor already close to the full WZ tensor, offering a far more effective starting point for fitting than an unguided approach.

\subsection{VFF Model Parameterization}
\label{subsec:fitting} 

To determine the force constants and $S$ parameter of the VFF potential, Eq.~(\ref{eqn:MartinPot}), we construct a workflow that employs the analytical expressions developed for ZB structures and exploits the similarities of the elastic tensor of WZ and [111]-oriented ZB systems. Moreover, we have performed DFT calculations to ensure consistent target parameter sets for the VFF parameterization. As test systems we focus on III-N materials, which preferentially crystallize in the WZ phase and are of high technological relevance~\cite{HumphreysSolidStateLighting}. Moreover, and in contrast to other VFF parameter extraction schemes, we focus not only on structural but also electronic properties to establish and optimize our VFF model. To do so, the VFF relaxed atomic positions are used directly as input to DFT calculations to determine the electronic structure and, in particular, the band gap of the different materials. The band gap data are compared with results from DFT-relaxed atomic positions. We focus on the band gap as it is a critical parameter for optoelectronic devices, where III-N materials are widely used. Our VFF model parameterization can be divided into three main steps, which are schematically depicted in Fig.~\ref{fig:workflow}.

\emph{\textbf{Step 1: Elastic Constants}}
To construct the WZ VFF model from Eq.~(\ref{eqn:MartinPot}), we start with the procedure outlined already above. This means we use the analytical expressions in 
Appendix~\ref{sec:appendix}, Eqs.~(\ref{eqn:fc_ktheta})~-~(\ref{eqn:Sparam}), to obtain initial values for force constants. These equations rely on the elastic 
constants of the [001]-oriented ZB system, where the three independent constants $C^{\text{ZB}}_{11}, C^{\text{ZB}}_{12}, C^{\text{ZB}}_{44}$ also define the 
elastic constants $C'_{ij}$ of the [111]-oriented ZB system, Eqs.~(\ref{eqn:001ZBto111ZB_start})~-~(\ref{eqn:001ZBto111ZB_end}). Since the elastic tensor for [111]-oriented ZB closely resembles that of WZ, Eq.~(\ref{matrix:WZ}), varying $C^{\text{ZB}}_{11}, C^{\text{ZB}}_{12}, C^{\text{ZB}}_{44}$ yields a [111]-ZB tensor that approximates the WZ tensor. To achieve this, a least-square fitting procedure, based on Nelder-Mead Simplex method~\cite{NeMe1965} and implemented in the programming language \textsc{julia}~\cite{Julia2017} using the \textsc{Optim.jl} package~\cite{OptimJL2018}, is employed. Thus we determine $\tilde{C}_{11}^\text{ZB}$, $\tilde{C}_{12}^\text{ZB}$ and $\tilde{C}_{44}^\text{ZB}$ constants that give a [111]-ZB tensor mimicking the desired WZ tensor: 
\begin{equation}
D_\text{ela}=\sum_{ij} w_{ij}\left(C^\text{WZ}_{ij}-C^{'}_{ij}(C^\text{ZB}_{11},C^\text{ZB}_{12},C^\text{ZB}_{44})\right)^{2}\,\, .
\label{eq:LS_1}
\end{equation}
The factor $w_{ij}$ allows assigning different weights to different elastic constants, i.e., certain elastic constants can be given a higher priority in the fitting procedure. However, careful attention must be paid to the stability of the crystal against sub-lattice displacements. As demonstrated by Tanner \emph{et al.}~\cite{TaCa2019}, covalent materials must satisfy $C_{11}-C_{12} > C_{44}$. This condition is commonly expressed through the anisotropy parameter $A$:
\begin{equation}
A =\frac{2C_{44}}{C_{11}-C_{12}} <2 \,\, .
\end{equation}
Thus, in the case of covalent materials, this constraint is essential when using the least-square fitting procedure based on Eq.~(\ref{eq:LS_1}). For highly ionic materials, such as III-N materials, this condition is not sufficient~\cite{TaCa2019}. In this case, the Coulomb interaction parameter, $S$, see Eq.~(\ref{eqn:Sparam}), must remain positive. We include these constraints in our fitting procedure of the elastic tensor when minimizing $D_\text{ela}$.

\emph{\textbf{Step 2: Force constants}}
The  elastic constants $\tilde{C}_{11}^\text{ZB}, \tilde{C}_{12}^\text{ZB}, \tilde{C}_{44}^\text{ZB}$ determined in Step 1 yield an elastic tensor of [111]-oriented ZB that approximates WZ, and the equations in Appendix~\ref{sec:appendix} determine the force constants and $S$ required for Eq.~(\ref{eqn:MartinPot}) analytically.  Here, in addition to $\tilde{C}_{11}^\text{ZB}$, $\tilde{C}_{12}^\text{ZB}$ and $\tilde{C}_{44}^\text{ZB}$, the internal strain parameter (Kleinman parameter), $\zeta$, the equilibrium lattice constant of ZB, $a_{0}$, and the inner elastic constant, $E_{11}$, are inputs. More detail and discussion of $\zeta$ and $E_{11}$ can be found in Refs.~\cite{TaCa2019,CaSc2013}. Given that the nearest neighbor environment is almost identical for WZ and [111]-oriented ZB, we use and determine $\zeta$ and $E_{11}$ from ZB values.  For the equilibrium lattice constant, defined by $a_{0}= \frac{\sqrt{3}}{4} r^{0}_{1}$, where $r^0_{1}$ is the anion-cation bond length in ZB, we use an average of the two WZ bond lengths $r^0_{1}$ and $r^0_{2}$ (see Fig.~\ref{fig:structures}) to obtain an initial value for $a_0$. 

Equipped with this initial guess for the force constants and the $S$ parameter, we proceed as follows to establish a VFF model for WZ. In a first step, the VFF model in Eq.~(\ref{eqn:MartinPot}) is modified to treat the WZ specific aspect of the nearest neighbor bond lengths asymmetry. Different approaches have been proposed in the literature to capture this effect~\cite{MaZu1998,GrNe2001,CaNi2010}, and we introduce the WZ bond length asymmetry through the Coulombic linear repulsion term, $V_\text{rep}$. Thus, for our WZ VFF model, $V_\text{rep}$ in Eq.~(\ref{eqn:MartinPot}) is split into two contributions accounting for the two different nearest neighbor bond lengths: 
\begin{align}
V_\text{rep}=
    \label{eqn:WZPot}
    \frac{1}{2} \Bigg[ &
    \sum_{j \in r_1}^{nn} 
        \frac{1}{4} \alpha_{M} S 
        \frac{e^2}{4 \pi \epsilon_0 {r_{1}^{0}}^{2}} 
        \left(r_{ij} - r_{1}^0\right)
    \nonumber \\[4pt]
    &\quad + 
    \sum_{j \in r_2}^{nn} 
        \frac{1}{4} \alpha_{M} S 
        \frac{e^2}{4 \pi \epsilon_0 {r_{2}^{0}}^{2}} 
        \left(r_{ij} - r_{2}^0\right)
    \Bigg].
\end{align}
The benefit of employing this extension to the model is that (i) it does not introduce any new or additional free parameters and (ii) it is still consistent with the potential, Eq.~(\ref{eqn:MartinPot}), that describes a ZB crystal structure, if the bond length asymmetry is absent. The WZ VFF is implemented in the software package \textsc{gulp} v.6.1.2~\cite{GULP1997, GULP2003},  
which allows us to calculate the elastic tensor of the chosen materials. To do so, the crystal structure is relaxed to its minimum energy configuration. This involves a geometry optimization where the positions of the atoms and the lattice cell parameters are adjusted until the forces on the atoms and the stress on the cell are minimized. To obtain the elastic tensor, the optimized structure and the corresponding energy are used to calculate the second-order derivatives of the energy with respect to strain. This second-derivative matrix is the elastic tensor of, in our case, a WZ unit cell. 

Fitting the different force constants $k_{r}$, $k_{\theta}$, $k_{rr}$, $k_{r\theta}$, and $S$ to the elastic properties
of the WZ material, namely $C_{11}$, $C_{12}$, $C_{13}$, $C_{33}$ and $C_{44}$, results in a good quantitative
description of these properties, as demonstrated below. Again we employ the Nelder-Mead Simplex method~\cite{NeMe1965} to establish a least square fitting procedure:
\begin{equation}
    D^\text{WZ}_\text{ela} = \sum_{i.j}^{N_{\alpha}} w_{ij} (C^\text{WZ,T}_{ij} - C^\text{WZ,P}_{ij}(k_{r}, k_{\theta}, k_{rr}, k_{r\theta}, S))^2\,\, .
    \label{eqn:sumsquares}
\end{equation}
Here, $C_{ij}^\text{WZ,T}$ are the target observables to which the VFF model is fitted, and $C_{ij}^\text{WZ,P}$ denotes the predicted elastic constants by the VFF model. Equation~(\ref{eqn:sumsquares}) allows us to include a weighting factor $w_i$. Here, we equally weigh all elastic constants. The cost function, Eq.~(\ref{eqn:sumsquares}), is minimized to establish our VFF parameter set for WZ. In the final stage of our model development, we target the bond-length asymmetry with greater precision, a crucial factor governing the electric polarization fields in WZ \mbox{III-N} systems~\cite{CaSc2013}. Moreover, we optimize to the material band gap, which is usually not considered in such structural models.

\emph{\textbf{Step 3: Bond asymmetry and band gap:}}
In Step 2, we determined the VFF force constants and the $S$ parameter by fitting to the WZ elastic tensor. This is the general procedure found in the literature for establishing VFF models~\cite{CaNi2010, ShTa2021, NEMO2007}. We now go one step further in our model parametrization, paying close attention to whether the model captures (i) the bond asymmetry of the WZ system and (ii) the electronic structure, in our case the band gap. While (i) can be directly assessed from  our \textsc{gulp} calculations, for (ii) an electronic structure model is required. For this, we use the DFT framework that is employed to determine the reference elastic constants. The band gap is calculated with DFT using VFF‑relaxed atomic positions (Step 2) and compared to a full DFT calculation, where the atomic positions are relaxed entirely by first‑principles methods.

Simultaneous agreement in bond‑length asymmetry and band gap is efficiently achieved by tuning the $S$ parameter, which governs the Coulombic terms in the VFF model, including the linear repulsion term, Eq.~(\ref{eqn:WZPot}). Increasing (decreasing) $S$ enhances (reduces) the bond‑length asymmetry.
In this optimization step, the $S$ term is adjusted, and then fixed, and the force constants are refit using the same procedure as in Step 2 but without adjusting $S$ in Eq.~(\ref{eqn:sumsquares}). This is repeated, where $S$ is altered and the force constants are refit, until the bond lengths and band gap are best described 
without significantly compromising on the elastic constants. 

This optimization step entails some degree of freedom in terms of what is deemed a good fit. We have set an upper limit of $\pm10\%$ deviation for all observables, i.e., elastic constants, bond lengths and band gap.  
We will show that the observables predicted by our VFF model are well within this margin; the largest deviation found is approximately $7\%$ with many observables exhibiting errors close to $0\%$.

Importantly, the model not only captures the expected deviations of $c_0/a_0$ ratio and internal parameter $u$ from their ideal values ($\sqrt{8/3}$ and $3/8$), but also reproduces the DFT results with high fidelity. We also observe that electronic structure properties, such as the crystal‑field splitting energies, are reasonably well reproduced when using the VFF‑relaxed atomic positions in DFT calculations. This presents a further validation of the model, given that these quantities have not been included in the fitting procedure established in the workflow.

\section{Results: VFF model for {InN}, {GaN}, {AlN}, {BN} and boron containing III-N alloys}
\label{sec:boron} 

In this section we use the WZ III-N binaries InN, GaN, AlN and BN as examples for establishing a VFF model based on the procedure outlined in the previous section. Introducing WZ BN into III-N alloys has garnered significant interest to tailor and enhance the efficiencies of III-N optoelectronic devices~\cite{SaMs2021}. This necessitates a deeper understanding of the fundamental properties of these boron-containing alloys. We present a detailed comparison between results from our VFF model for (B,Ga)N and (B,In,Ga)N alloys and full DFT calculations, focusing on atomic positions and band-gap predictions. This direct benchmarking of band gaps goes beyond conventional methodologies widely employed in the literature. To do so, we outline in the following section the DFT approach applied to obtain structural and electronic properties of III-N systems required for the development of the VFF model and connected benchmarking.

\begin{table*}
\begin{ruledtabular}
\begin{tabular}{ccllllllll}
     & & \multicolumn{1}{c}{$C_{11}$ (GPa)} & $C_{33}$ (GPa) & $C_{12}$ (GPa) & $C_{13}$ (GPa) & $C_{44}$ (GPa) & $r^0_{1}$ (\AA) & $r^0_{2}$ (\AA) & $E_g$ (eV)\\
    \hline 
    \multirow{3}{*}{InN} & DFT & 195.2 & 211.8 & 98.4 & 71.4 & 45.15 & 2.185 & 2.191 & 0.69 \\
    & VFF & 201.7 (3.3\%) & 206.3 (-2.6\%) & 92.8 (-5.7\%) & 75.9 (6.3\%) & 47.5 (5.2\%) & 2.187 (0.1\%) & 2.191 (0.0\%) & 0.68 (-1.5\%) \\
    & Lit. & 233.8$^{a}$  & 238.3$^{a}$ & 110$^{a}$ & 91.6$^{a}$ & 55.4$^{a}$ & - & - & 0.68$^{c}$ \\
    \hline
    \multirow{3}{*}{GaN} & DFT & 323.6 & 366.1 & 116.8 & 71.7 & 91.8 & 1.967 & 1.975 & 3.51 \\
    & VFF & 330.4 (2.1\%) & 359.1 (-1.9\%) & 111.3 (-4.7\%) & 76.7 (7.0\%) & 96.2 (4.8\%) & 1.971 (0.2\%) & 1.973 (-0.1\%) & 3.51 (0.0\%)\\
    & Lit. & 368.6$^{a}$ & 406.2$^{a}$ & 131.6$^{a}$ & 95.7$^{a}$ & 101.7$^{a}$ & - & - & 3.24$^{c}$ \\
    \hline
    \multirow{3}{*}{AlN} & DFT & 382.1 & 354.5 & 126.5 & 97.3 & 109.9 & 1.902 & 1.912 & 6.09\\
    & VFF & 376.5 (-1.5\%) & 355.8 (0.4\%) & 127.5 (0.8\%) & 102.5 (5.4\%) & 109.2 (-0.6\%) & 1.903 (0.1\%) & 1.919 (0.4\%) & 6.01 (-1.3\%)\\
    & Lit. & 410.2$^{a}$ & 385.0$^{a}$ & 142.4$^{a}$ & 110.1$^{a}$ & 122.9$^{a}$ & - & - & 5.64$^{c}$ \\\hline
    \multirow{3}{*}{BN}  & DFT & 921.8 & 1006.3 & 129.9 & 56.4 & 327.5 & 1.567 & 1.583 & 6.70 \\ 
    & VFF & 921.9 (0.0\%) & 1003.3 (-0.3\%) & 133.8 (3.0\%) & 55.6 (-1.4\%) & 334.7 (2.2\%) & 1.574 (0.5\%) & 1.579 (-0.3\%) & 6.85 (2.2\%)\\ 
    & Lit. & 1016$^{b}$ & 1113$^{b}$ & 144$^{b}$ & 64$^{b}$ & 361$^{b}$ & - & - & 6.71$^{b}$\\
\end{tabular}
\end{ruledtabular}
\caption{Elastic constants, bond lengths and band gaps for wurtzite InN, GaN, AlN, and BN obtained from our density functional theory (DFT) and valence force field (VFF) model calculations. The percentage deviation between VFF and DFT is given in brackets. We provide literature values (Lit.) taken from HSE06-DFT (apart from Ref.~\cite{Sheerin2022} which is HSE-DFT with an adjusted exact exchange mixing parameter) for comparison for all parameters except the bond lengths, which are not usually provided in the literature. \\
$^{a}$Reference~\cite{Caro2012} \\
$^{b}$Reference~\cite{Sheerin2022} \\
$^{c}$Reference~\cite{YaRi2011}}

\label{tab:elastic_constants_bond_gap}
\end{table*}

\subsection{DFT Calculations}
\label{sec:DFT}

All DFT calculations have been performed using the Vienna ab initio Simulation Package (\textsc{VASP}) v5.4 \cite{VASP}. In the literature, III-N semiconductors have often been targeted by Heyd, Scuseria, and Ernzerhof (HSE) hybrid functional density functional theory (HSE-DFT) to bypass band gap problems~\cite{xiao11, perdew17, lany08}.  
However, HSE-DFT is computationally expensive and the larger supercells that are required for low alloy concentrations present challenges, especially if calculations have to be repeated several times to sample the impact of the alloy microstructure on the electronic and optical properties~\cite{HSEDFT,Krukau2006_HSE, BrOR2024}.

As we aim to establish a VFF model for both binary systems and larger alloy supercells with varying (random) alloy configurations, it is essential to employ a consistent DFT approach to provide benchmarks and validate the model. Moreover, the DFT calculations should give a good description of the band gap of the materials. Thus, we employ the meta-generalized gradient approximation (meta-GGA), which is significantly less computationally 
expensive compared to HSE-DFT but produces commensurate results in terms of band gaps \cite{Borlido2020}. More specifically, we employ the modified Becke–Johnson (mBJ) meta-GGA functional for all band gap calculations~\cite{mBJ, Tran2009}. Given that the mBJ meta-GGA functional is a potential only method, these functionals do not allow for self-consistent calculations with respect to the total energy. Therefore, for geometry and lattice optimizations, we use the generalized gradient approximation (GGA) based on the Perdew–Burke–Ernzerhof (PBE)~\cite{PBE} ansatz for the exchange correlation functional. The GGA approach is chosen because it more accurately treats systems with non‑homogeneous electron densities than the local density approximation, making it better suited for modeling alloy disorder~\cite{koch01-4}. 

For all our bulk binary systems we use a $\Gamma$-centered $6\times6\times4$ Monkhorst-pack $k$-point mesh and a Gaussian 
smearing of $\sigma$ = 0.1 eV; the plane wave cut-off energy is 600 eV. We include the Ga and In $d$-electrons as valence electrons. To find the equilibrium parameters, the pressure on the unit cells is minimized through the strain stress relationship~\cite{CaSc2013_2}. The elastic constants are also determined through the strain stress method. Details on band structures, mBJ parameters, benchmarks against HSE-DFT and the calculation of elastic constants can be found in our previous work \cite{CaSc2013_2, NiSh2023}.
The binary bulk calculations allow us to produce the data required to parameterize the VFF model.

We examine the predictive power of the VFF model by focusing on (B,Ga)N and (B,In,Ga)N alloys, which have attracted interest for optoelectronic device application~\cite{GaOr2010, LoKi2017,SaMs2021}. 
To establish a DFT benchmark for our VFF model, we constructed $(3\times 3\times 3)$ supercells containing 108 atoms. These supercells are sufficiently large to model experimentally relevant boron concentrations as low as 2\%, by replacing 1 of 54 group III-atoms with B. We maintain a consistent $k$-point density  between supercell and binary (unit cell) calculations by employing a $2\times2\times1$ $\Gamma$-centered Monkhorst-pack $k$-point mesh for the supercell. All other settings are kept identical to those used for the binary systems.

\subsection{VFF model for WZ InN, GaN, AlN and BN}
\label{subsec:BN_fitting}

In this section, we present the VFF parameters for BN, AlN, GaN and InN, obtained following the workflow outline in Sec.~\ref{subsec:fitting}. 
We use the DFT methods described above to provide the input required for VFF parameterization, namely elastic constants, bond lengths and band gaps. 
The relevant DFT data are summarized in Table~\ref{tab:elastic_constants_bond_gap}. Our PBE-DFT elastic constants, $C_{ij}$, are smaller than literature HSE-DFT values, which is a well-know effect as PBE-DFT tends to underestimate binding energies~\cite{RaMo2015}. However, as outlined above, we have selected PBE-DFT to provide a consistent DFT framework that can target larger supercells efficiently  from the perspective of computational cost. Moreover, as Table~\ref{tab:elastic_constants_bond_gap} shows, our mBJ-DFT band gaps, obtained with underlying PBE-DFT for geometry optimization, are in good agreement with literature HSE-DFT data; we note that the above data is based on HSE06 results and further improvements in the band gap can be achieved by adjusting the exact exchange parameter~\cite{MoMi2011}. Thus, the above DFT framework using PBE and mBJ functionals is sufficient to establish a reliable VFF model for III-N materials. 

The PBE-DFT data are then used in the VFF parameter extraction workflow discussed in Sec.~\ref{subsec:fitting}. The resulting force constants and $S$ parameter underlying the VFF model are given in Table~\ref{tab:force_constants} for each binary material. Table~\ref{tab:elastic_constants_bond_gap} summarizes the VFF model output for structural properties and band gap values. Our VFF model delivers accuracy exceeding 92\% relative to DFT: elastic constants, bond lengths, and band gaps are reproduced within 7\%.

\begin{table}[b!]
    \centering
    \begin{ruledtabular}
    \begin{tabular}{c c c c c c}
               & $k_{r}$ & $k_{\theta}$ & $k_{r r}$ & $k_{r \theta}$ & $S$ \\
         Units &  eV \r{A}$^{-2}$ & eV rad$^{-2}$ & eV  \r{A}$^{-2}$ & eV  \r{A}$^{-1}$  rad$^{-1}$ &  \\
         \hline \\[0.1cm]
         InN   & 10.582 & 1.526 & 0.884 & 0.314 & 0.810\\
         GaN & 14.741 & 2.506 & 1.765 & 0.097 & 0.422 \\
         AlN & 18.488 & 1.857 & 1.385 & 0.232 & 2.116 \\
         BN & 17.584 & 4.909 & 4.200 & 1.297 & 0.040 \\
    \end{tabular}
\caption{Force constant for the valence force field potential for binary WZ III-N systems.} 
    \label{tab:force_constants}
    \end{ruledtabular}
\end{table}

\begin{table*}
    \centering
    \begin{ruledtabular}
    \begin{tabular}{cllllll}
        & & \textit{a}\textsubscript{0} (\AA) & \textit{c}\textsubscript{0} (\AA) & \textit{u} & \textit{$c_0/a_0$} & $\Delta_\text{CF}$ (meV) \\
        \hline
        \multirow{3}{*}{InN} & DFT & 3.581 & 5.795 & 0.378 & 1.618 & 29 \\
                             & VFF & 3.584 (0.1\%) & 5.798 (0.1\%) & 0.378 (0.0\%) & 1.618 (0.0\%) & 29 (0.0\%) \\
                             & Lit & 3.548$^{a}$ & 5.751$^{a}$ & 0.380$^{a}$ & 1.621$^{a}$ & 19-66$^{a}$ \\
        \hline
        \multirow{3}{*}{GaN} & DFT & 3.218 & 5.244 & 0.377 & 1.630 & 22 \\
                             & VFF & 3.222 (0.1\%) & 5.243 (0.0\%) & 0.376 (-0.3\%) & 1.627 (-0.2\%) & 41 (86.4\%) \\
                             & Lit & 3.182$^{a}$ & 5.173$^{a}$ & 0.377$^{a}$ & 1.626$^{a}$ & 9-38$^{a}$ \\
        \hline
        \multirow{3}{*}{AlN} & DFT & 3.127 & 5.017 & 0.381 & 1.604 & -180 \\
                             & VFF & 3.135 (0.3\%) & 5.015 (0.0\%) & 0.383 (0.5\%) & 1.600 (-0.3\%) & -242 (34.4\%) \\
                             & Lit & 3.102$^{a}$ & 4.971$^{a}$ & 0.382$^{a}$ & 1.603$^{a}$ & (-169)-(-295)$^{a,c}$ \\
        \hline
        \multirow{3}{*}{BN}  & DFT & 2.555 & 4.224 & 0.375 & 1.654 & 246 \\
                             & VFF & 2.572 (0.7\%) & 4.199 (-0.6\%) & 0.376 (0.3\%) & 1.633 (-1.3\%) & 118 (-52.0\%) \\
                             & Lit & 2.534$^{b}$ & 4.189$^{b}$ & 0.375$^{b}$ & 1.653$^{b}$ &  $\approx 250^{d}$ \\
    \end{tabular}
  \caption{Lattice parameters, $a_0$ and $c_0$, and internal $u$ parameter along with the crystal field splitting energy, $\Delta_\text{CF}$, for WZ binary nitrides. The data are given the same format as Table~\ref{tab:elastic_constants_bond_gap}. Once again we provide literature values (Lit.) taken from HSE06-DFT (apart from Ref.~\cite{Sheerin2022} and Ref.~\cite{NiSh2023} which are HSE-DFT) for comparison for all parameters, except for $\Delta_\text{CF}$ where the literature values are from a range of calculated and experimental results. \\
    $^{a}$Reference~\cite{YaRi2011} \\
    $^{b}$Reference~\cite{Sheerin2022} \\
    $^{c}$Reference~\cite{VuMe2003} \\
    $^{d}$Reference~\cite{NiSh2023}}
    \label{tab:binary_bls}
    \end{ruledtabular}
\end{table*}

The above comparison focuses on properties that are directly underlying the parameterization of the VFF model. In a second step, we now employ the model to 
\emph{predict} data that have not been used to train it. Here, we target structural as well as electronic band structure properties. As such we have used the bulk 
lattice constants, $a_0$ and $c_0$, the internal parameter, $u$, of the WZ lattice~\cite{Caro2012} and the crystal field splitting energy, $\Delta_\text{cf}$, as the target 
properties for this benchmark. Capturing the magnitude but in particular the sign of $\Delta_\text{cf}$ plays an important role for, e.g., light extraction 
efficiencies in deep ultraviolet light emitters based on (Al,Ga)N alloys~\cite{FiOD2025}. The deviation of $u$ from its ideal value, $u^\text{id}=3/8=0.375$, is 
tightly linked to the large spontaneous electric polarization found in III-N systems~\cite{DrJo2016}. Therefore, it is of key importance to capture the $u$ value accurately  when using methods such as local polarization theory to determine built-in electric fields in electronic structure calculations of III-N 
heterostructures~\cite{CaSc2013,ScCa2015}. 

In Table~\ref{tab:binary_bls}, we present the DFT and predicted VFF data, without any further fitting. Looking at the $u$ parameter first, our VFF model predicts values in excellent agreement with the DFT data, which already reflect the expected deviation from $u^\text{id}$ to be compatible with large spontaneous polarization values. Given that $u$ departs from its ideal value, the lattice constant ratio correspondingly deviates from the ideal WZ value of $\left(c_0/a_0\right)^\text{id}=\sqrt{8/3}\approx1.633$. Again we find excellent agreement between DFT and VFF data, both in terms of absolute $a_0$ and $c_0$ values and for the $c_0/a_0$ ratio, highlighting further that characteristics of realistic III-N materials are captured by the VFF potential. 

Turning to the crystal field splitting energy, our VFF model captures the correct trends in $\Delta_\text{cf}$, predicting $\Delta_\text{cf}>0$ for InN, GaN, and BN and $\Delta_\text{cf}<0$ for AlN. While the magnitudes for BN, AlN, and GaN differ from our DFT reference, InN is reproduced exactly. Importantly, crystal field splitting energies themselves are known to exhibit significant variations in the literature, and the values predicted by our VFF model fall well within the reported ranges, except for BN where limited information exists, see Table~\ref{tab:binary_bls}. Thus, these differences do not undermine the reliability of our model, especially since it is a tool for structural relaxations. For the crystal field splitting energies, not only geometrical but also electronic (charge density) features may play a role. Depending on the material system, for example, (Al,Ga)N, one may consider incorporating $\Delta_\text{cf}$ into the fitting procedure (see Sec.~\ref{subsec:fitting}), though this lies beyond the scope of the present study. In fact, $\Delta_\text{cf}$ can also be further refined in empirical electronic structure models based on using the equilibrium bond lengths of our model in band structure fitting~\cite{KoSa83,Joga98,CaSc2013}.

Overall, our VFF model serves as a robust tool for structural relaxations. It not only reproduces the parameters to which it is fitted, such as the elastic tensor and bond lengths, but also reliably predicts independent observables like the internal parameter $u$ and the $c_0/a_0$ ratio. The next critical step, however, is extending this capability to capture relaxed atomic positions in disordered systems, a challenge we address in the following section.

\subsection{Boron containing III-N alloys}
\label{subsec:alloys} 

In this subsection, we focus our attention on boron containing III-N alloys and validate our VFF potential for use as a tool for structural optimization. First, we target ternary (B,Ga)N alloys before turning to the quaternary alloy (B,In,Ga)N. As discussed above, these alloys  are of interest for optoelectronic device applications~\cite{GaOr2010, LoKi2017,SaMs2021}. The relaxed atomic positions obtained from the VFF model are compared with positions predicted by DFT calculations performed on the same supercells. Moreover, the VFF optimized structures are directly used, without further relaxation, in band gap calculations via mBJ-DFT, as in Sec.~\ref{subsec:BN_fitting}. The resulting band gaps are compared to full DFT calculations, by which the relaxed atomic atomic positions are determined by DFT.

\subsubsection{(B,Ga)N alloys}
\label{subsec:BGaN}

In the following, we analyze the VFF capabilities for predicting structural properties of (B,Ga)N alloys and connected effects on the band gap value. Special attention is paid to the impact of the alloy microstructure. As already mentioned, experimentally small amounts of WZ BN can be incorporated into GaN, typically in the range of $<10$\% BN to achieve good crystalline quality~\cite{GaOr2010,LiLi2020, TrLi2020,SaMs2021}.  
Based on these experimental constraints, we investigate properties of (B,Ga)N alloys between $\approx2$\% and $\approx7$\%. To be able to study these low compositions, larger supercells, which can also be treated efficiently in DFT, must be used. Our $3\times3\times3$ supercells with 108 atoms allow us to access these experimentally relevant concentrations. Here, replacing 1 of the 54 Ga atoms with a B atom is equivalent to $\approx2$\% BN in GaN. In the following, we label the systems and compositions studied in terms of the number of B, Ga and N atoms in the supercell. Thus, a (B,Ga)N system with $\approx2\%$ BN is denoted by B$_1$Ga$_{53}$N$_{54}$. 

We investigate the impact of composition and alloy microstructure on structural and electronic properties by proceeding as follows: Firstly, we target the low BN content regime with $\approx$ 2\%. Secondly, we consider an intermediate BN concentration with $\approx$ 5\% BN, whereby 3 Ga atoms in the supercell are replaced by 3 B atoms. Since the alloy microstructure can significantly affect properties such as the band gap~\cite{NiSh2023}, we construct three random alloy configurations of B atoms in the GaN supercell by substituting Ga with B. These random alloy structures are labeled RA I, RA II, and RA III. 
Finally, we investigate a system with $\approx$ 7\%, which is toward the higher end of what is experimentally viable without compromising on material 
quality. In this case the supercell contains 50 Ga and 4 B atoms (B$_4$Ga$_{50}$N$_{54}$). Building on literature reports of B atom clustering~\cite{SaMs2021} and our prior work showing its impact on the electronic structure~\cite{NiSh2023}, we examine an extreme case where four B atoms share a common N atom. This forms a BN tetrahedron within the GaN matrix, and gives insight into whether our model can capture these extreme clustering and strain cases.

\begin{figure*}
    \centering
    \includegraphics[width=1\linewidth]{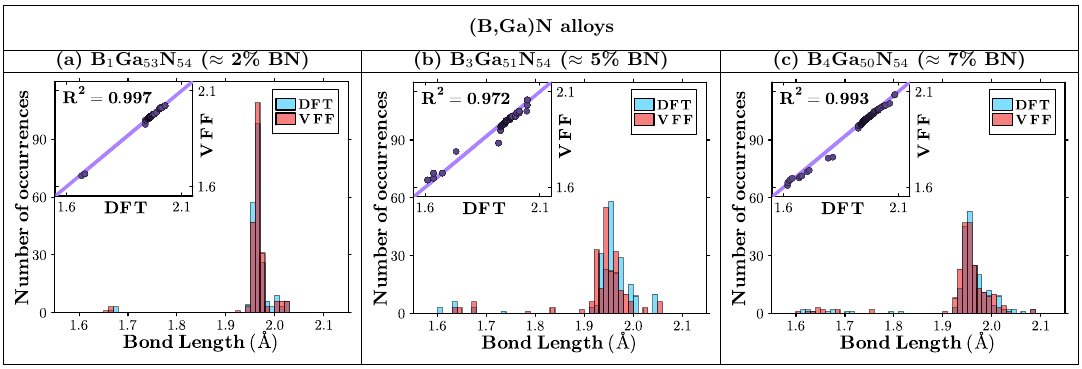} 
    \caption{Comparison of bond length distributions from density functional theory (DFT) and valence force field model (VFF) for (B,Ga)N alloys, with a histogram bin value of 0.01\,\AA. Inset: a quantile-quantile plot comparing the distributions, where the solid line indicates perfect agreement between VFF and DFT. For the B$_{3}$Ga$_{51}$N$_{54}$ ($\approx$ 5\% BN) we have plotted results from the random alloy configuration labeled as RA II in the main text.}
    \label{fig:bl_distributions}
\end{figure*}

We apply the VFF potential, originally developed for binary systems, to (B,Ga)N as follows: Force constants describing two-body interactions, e.g. 
$k_r$, are set to the bulk values as they account for interactions between B-N or Ga-N atoms. For terms that involve multi-cation interactions we use 
a linear interpolation of the binary values. This includes the three-body term $k_\theta$, which involves, for example, a central N atom bonded with a 
Ga and a B atom. Finally, for the Coulombic terms, at cation sites the binary $S$ values are employed. For N atoms, which have varying numbers of Ga 
and B atoms as nearest neighbors, weighted averages of the $S$ parameter, based on Ga and B atom numbers, are set. This procedure ensures that the 
alloy is charge neutral and consequently stable. The above procedure is not dissimilar to establishing empirical tight-binding models for alloys, 
where hopping (interaction) matrix elements are set to their binary values and onsite energies are described by weighted 
averages~\cite{OReLi20021,BoKh2007,ScCa2015}. In general, more advanced parameterization strategies can be envisioned by explicitly calculating the 
elastic tensor of a disordered alloy system and then incorporating this tensor into the fitting procedure. Such an approach, however, is considerably 
more demanding: alloy disorder breaks the underlying crystal symmetry, thereby introducing additional elastic constants into the tensor. Moreover, 
meaningful averages require calculations over several distinct random alloy configurations. Here, the adopted simpler strategy reproduces, 
nevertheless, structural and electronic properties of (B,Ga)N alloys across the experimentally relevant composition range with high fidelity. This 
demonstrates that our approach is not only practical but also sufficient for capturing the essential physics of these alloys.

Figure~\ref{fig:bl_distributions} displays the bond length distributions obtained from VFF and DFT calculations for (a) B$_{1}$Ga$_{53}$N$_{54}$ ($\approx 2$\% 
BN), (b) an random alloy (RA) structure for the B$_{1}$Ga$_{53}$N$_{54}$ ($\approx$ 5\% BN) system and (c) the B$_{4}$Ga$_{50}$N$_{54}$ ($\approx$ 7\% BN) case with B atom clustering (tetrahedron of B around N atom). The insets show quantile-quantile (Q-Q) plots, which allow a further means for comparing the bond length distributions from VFF and DFT. In case of perfect agreement, all the displayed points will lie exactly on a straight diagonal line in a Q-Q plot, indicating that the two datasets being compared have identical 
distributions. Considering Fig.~\ref{fig:bl_distributions} (a), the histogram reveals one main peak around the bond lengths of unstrained GaN (\mbox{$r^\text{GaN, DFT}_1=1.967$ \AA}, \mbox{$r^\text{GaN, DFT}_2=1.975$ \AA}). This may not be surprising as only a single B atom in a GaN matrix is investigated. No peak is observed at the unstrained BN bond lengths ($r^{\text{BN},\text{DFT}}_1 = 1.567 \,\text{\AA}, \; r^{\text{BN},\text{DFT}}_2 = 1.583 \,\text{\AA}$). This stems from the fact that in order to minimize the elastic energy of the system, and given that $r^{\text{GaN},\text{DFT}}_i > r^{\text{BN},\text{DFT}}_i$, the bonds surrounding the B atom are stretched. In contrast, the Ga–N bonds in the vicinity of B atoms experience both tensile and compressive distortions. 
The VFF models reproduces the trends seen with DFT accurately. The Q-Q plot confirms this, as all data points almost perfectly lie on a straight diagonal line. We also calculated the $R^2$ value, which provides a quantitative summary of how well the DFT data are replicated by our VFF model. In general, an $R^2=1$ means perfect agreement, while $R^2=0$ indicates no agreement at all. For the B$_{1}$Ga$_{53}$N$_{54}$ system we find $R^2=0.997$ reflecting and confirming that the VFF model accurately captures the bond length distribution compared to DFT. In general, Figure~\ref{fig:bl_distributions} 
(a) gives microscopic information of local changes in the atomic positions. We complement this by summarizing macroscopic features, i.e., size of the fully relaxed  supercell and the band gap in Table~\ref{tab:alloy_outputs}. The lattice constant of the supercell, $a$ and $c$, predicted by the VFF model are in excellent agreement with the DFT data. The same is true for the band gap $E_g$, indicating that even though there may be slight differences in the local atomic arrangements, these are of secondary importance for the band gap.

\begin{figure*}[t!]
    \centering
    \includegraphics[width=1\linewidth]{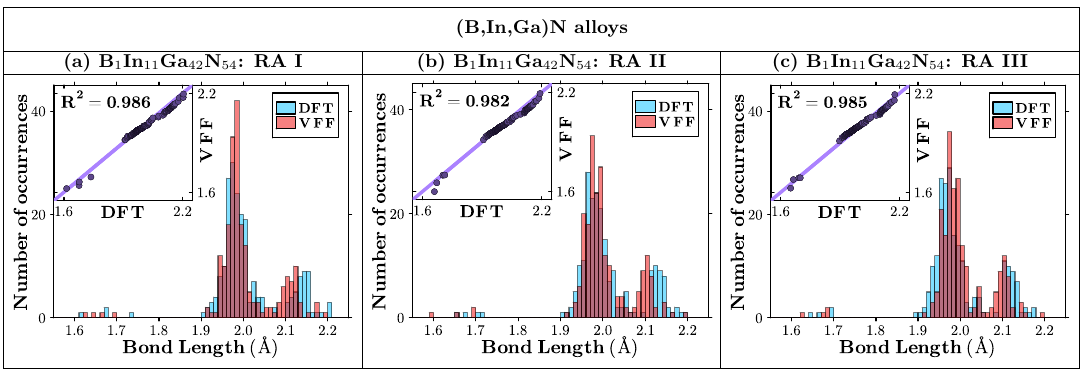} 
    \caption{Comparison of bond length distributions from density functional theory (DFT) and and valence force field (VFF) model for (B,In,Ga)N 108 atom supercells. The histogram bin values are 0.01\,\AA. Subplot: a quantile-quantile plot comparing the distributions, where the solid line indicates perfect agreement between VFF and DFT.}
    \label{fig:bl_distributions_BInGaN}
\end{figure*}

With reference to the random alloy case for B$_{3}$Ga$_{51}$N$_{54}$ ($\approx$ 5\% BN), Fig.~\ref{fig:bl_distributions} (b) displays the histogram for one of the 
three random alloy (RA) structures considered, which we denote as RA II henceforth. 
Overall, the histogram for RA II has a large peak around the equilibrium bond lengths of GaN, thus exhibiting a similar behavior when compared to B$_{1}$Ga$_{53}$N$_{54}$ ($\approx$ 2\% BN), Fig.~\ref{fig:bl_distributions} (a). However, the peak height is reduced in comparison to B$_{1}$Ga$_{53}$N$_{54}$, due to the increased number of B atoms in the system which cause a larger perturbation, leading to 
fewer unstrained Ga-N distances. Consequently, this also leads to an overall wider distribution of bond lengths values. The same trends are observed for RA I and RA III, which are summarized in Appendix~\ref{sec:appendixB}. Looking at the Q-Q plot for RA II, inset in Fig.~\ref{fig:bl_distributions} (b), we find good agreement between the bond length distribution predicted by the VFF model and the one obtained from DFT; this is also confirmed by a reasonably high $R^2=0.972$ value.
The lattice constants of the supercell after relaxation along with the band gap data are given in Table~\ref{tab:alloy_outputs}. This shows that our VFF optimization of RA II captures 
the structural relaxation well, which is further confirmed by the strong agreement between the band gap obtained with DFT using the VFF-relaxed atomic positions in  comparison to the full DFT optimization. 
This agreement is even better for the other two RA structures, RA I and III, see Table~\ref{tab:alloy_outputs}. 
The band gaps obtained for these three structures indicate that random alloy 
fluctuations ath this BN content, in general, lead to a slight reduction in band gap (compared to $E^\text{GaN}_g=3.51$ eV, see also Table~\ref{tab:elastic_constants_bond_gap})
We note that BN has a larger band gap than GaN and thus, one may not intuitively expect a \textit{increase} in band gap. However, this result is also in agreement with our previous DFT studies on (B,Ga)N alloys~\cite{NiSh2023}.

\begin{table}[b!]
    \centering
    \begin{ruledtabular}
    \begin{tabular}{c c l l l l }  
& & \textit{a} (\AA) & \textit{c} (\AA) & $E_{g}$ (eV)\\ 
 \hline\hline 
 & \multicolumn{4}{c}{\textbf{B$_{1}$Ga$_{53}$N$_{54}$ ($\approx$ 2\% BN)}}\\\hline
 \multirow{2}{*}{Single} & DFT & 9.618 & 15.669 & 3.50\\
                        & VFF & 9.619 (0.0\%) & 15.656 (-0.1\%)& 3.49 (-0.3\%)\\\hline
 & \multicolumn{4}{c}{\textbf{B$_{3}$Ga$_{51}$N$_{54}$ ($\approx$ 5\% BN)}}\\\hline                        
\multirow{2}{*}{RA I} & DFT & 9.548 & 15.556 & 3.47\\
                        & VFF & 9.528 (-0.2\%) & 15.521 (-0.2\%) & 3.47 (0.0\%)\\\hline

\multirow{2}{*}{RA II} & DFT & 9.548 & 15.555 & 3.46\\
                        & VFF & 9.529 (-0.2\%) & 15.527 (-0.2\%) & 3.42 (-1.2\%)\\\hline
\multirow{2}{*}{RA III} & DFT & 9.545 & 15.552 & 3.49\\
       & VFF & 9.522 (-0.2\%) & 15.508 (-0.3\%) & 3.49 (0.0\%)\\\hline
 & \multicolumn{4}{c}{\textbf{B$_{4}$Ga$_{50}$N$_{54}$ ($\approx$ 7\% BN)}}\\\hline       
\multirow{2}{*}{Cluster} & DFT & 9.528 & 15.492 & 3.20\\
         & VFF & 9.472 (-0.6\%)& 15.445 (-0.3\%) & 3.42 (6.9\%)\\
\end{tabular}
    \caption{Comparison of valence force field (VFF) and density functional theory (DFT) structural and band gap values obtained from 108 atom supercell calculations for different alloy contents and microstructures: one B atom (Single), three random alloy configurations (RA I-III) and 4 B  atoms sharing a N atom (Cluster).} 
    \label{tab:alloy_outputs}
     \end{ruledtabular}
\end{table}

For the GaN system with $\approx$ 7\% BN (B$_{4}$Ga$_{50}$N$_{54}$) and B atom clustering, Fig.~\ref{fig:bl_distributions} (c), we find a behavior very similar to the random alloy case. The large $R^2$ = 0.986 value indicates that the VFF model predicts a bond length distribution in very good agreement with DFT. The macroscopic features given in Table~\ref{tab:alloy_outputs} reveal that the $a$ and $c$ lattice constants are well reproduced by the VFF model. The band gap reduction compared to GaN is not as pronounced using the VFF-releaxed atomic positions as in a full DFT  calculation. However, the reduction in band gap associated with B atom clustering in (B,Ga)N alloys is preserved when structural relaxation is performed using our VFF model.

\subsubsection{(B,In,Ga)N alloys}
\label{subsec:BInGaN}

Beyond (B,Ga)N alloys, other BN-containing III–N compounds have attracted interest for the active regions of heterostructures, such as (B,In,Ga)N QWs embedded within GaN barriers. (B,In,Ga)N alloys offer a means to engineer strain and polarization fields in heterostructures, enabling emission from (In,Ga)N-based devices to be shifted toward longer wavelengths, including the red spectral region~\cite{LoKi2017}. Therefore, we use (B,In,Ga)N alloys as another test system for our VFF model. This provides an additional validation through incorporating four different elemental species instead of three as in the ternary case of (B,Ga)N and thus considering an even more complex system. 

To construct the VFF model for (B,In,Ga)N alloys we employ the same procedure as in the (B,Ga)N case and use linear interpolations and weighted averages of the 
binary parameters given in Table~\ref{tab:force_constants}. We consider again 108 atom supercells and construct three different random alloy configurations by 
randomly replacing Ga by B atoms in the simulation cell. The three structures are labeled as RA I, RA II and RA III. 
In the following we investigate the alloy B$_{1}$In$_{11}$Ga$_{42}$N, corresponding to $\approx2$\% BN, $\approx20$\% InN, and $\approx78$\% GaN, with InN and GaN compositions typical of (In,Ga)N light emitters in the visible range~\cite{HumphreysSolidStateLighting}. Although the BN content considered is only $\approx2$\% (one B atom in a 108-atom supercell), the system remains a highly mismatched alloy, with the lattice constant of BN differing from that of InN by 30\% (see $a_0$, Table~\ref{tab:binary_bls}). Thus, when a N atom is shared between B, Ga and In atoms, large local strains can be expected that must be captured by the VFF model.    

\begin{table}[b!]
    \centering
    \begin{ruledtabular}
    \begin{tabular}{c c l l l }
    \multicolumn{5}{c}{\textbf{B$_{1}$In$_{11}$Ga$_{42}$N$_{54}$}}\\\hline
& & \textit{a} (\AA) & \textit{c} (\AA) & $E_{g}$ (eV)\\ 
 \hline 
\multirow{2}{*}{RA I} & DFT & 9.843 & 15.985 & 2.63\\
       & VFF & 9.802 (-0.4\%) & 15.938 (-0.3\%) & 2.65 (0.8\%)\\\hline
\multirow{2}{*}{RA II} & DFT & 9.840 & 16.006 & 2.66\\
       &  VFF & 9.809 (-0.3\%) & 15.951 (-0.3\%) & 2.70 (1.5\%)\\\hline
\multirow{2}{*}{RA III} & DFT & 9.782 & 15.904 & 2.74\\
       &   VFF & 9.810 (0.3\%) & 15.952 (0.3\%) & 2.70 (-1.5\%)\\
\end{tabular}
    \caption{Comparison of valence force field (VFF) and denisty functional theory (DFT) lattice parameters, $a$ and $c$ and band gaps, $E_g$, obtained from 108-atom supercells for B$_1$In$_{11}$Ga$_{42}$N$_{54}$ random (RA) alloys. RA I to RA III denote three different random alloy configurations in the supercell.}
    \label{tab:BInGaN_outputs}
     \end{ruledtabular}
\end{table}

Figure~\ref{fig:bl_distributions_BInGaN} displays the bond length distributions for B$_{1}$In$_{11}$Ga$_{42}$N$_{54}$ in the case of (a) RA I, (b) RA II and (c) RA III. Again, these figures include a histogram and the Q-Q plot as an inset to compare distributions obtained within the VFF model and DFT. In contrast to the (B,Ga)N systems presented in Fig.~\ref{fig:bl_distributions}, we find two main peaks in the bond lengths distribution: one related to GaN (around $1.95-2.00$\, \AA) and one related to InN (around $2.10-2.15$\,\AA). 
Moreover, both models predict bond lengths in the range of $1.60 - 1.70$ \AA, which we attribute to strained B-N bonds. In general, the VFF model captures the trends seen in the DFT calculations very well, which is further supported by (i) the Q-Q plots, with data points lining up well on the diagonal, and (ii) the very high $R^2$ values of $>0.98$ (see insets in Fig.~\ref{fig:bl_distributions_BInGaN}).

Table~\ref{tab:BInGaN_outputs} summarizes the relaxed lattice constants and band gaps obtained from our DFT calculations. The VFF-derived lattice constants show excellent agreement with DFT, and using VFF-relaxed atomic positions in mBJ-DFT yields band gaps for all three RA structures that closely match full DFT results. Taken together, these results establish our VFF potential as a powerful framework for determining relaxed atomic positions in the HMA (B,In,Ga)N, and thus provides an ideal foundation for large-scale electronic structure calculations of these complex alloys and connected heterostructures.

\section{\label{sec:conclusion} Conclusion}

We developed a valence force field (VFF) model for wurtzite (WZ) systems, building on the fully analytical framework established by Tanner \emph{et al.}~\cite{TaCa2019} for zinc blende (ZB) crystal structures. Overall, the WZ model is guided by analytical expressions for force and elastic constants, and by structural similarities between WZ and [111]-oriented ZB. Unlike widely employed approaches that fit only to the elastic tensor, our method also targets bond asymmetries present in WZ and, through combination with density functional theory (DFT) calculations, the band gap of the materials under consideration. Although the method is general and can be applied to any WZ material, we focus on III-N compounds, including WZ BN, and connected highly mismatched alloys (B,Ga)N and (B,In,Ga)N, respectively. These materials and alloys have attracted strong interest for optoelectronic device applications in recent years. 

We have performed DFT calculations within the generalized gradient approximation (GGA) and meta-GGA framework to provide the data required for parameterizing and training the VFF model. The resulting VFF potential accurately reproduces the elastic tensor, bond lengths (including their asymmetry), and band gaps of WZ InN, GaN, AlN, and BN, with an accuracy of above 92\%. It also predicts structural parameters that the potential has not been fitted to -- such as the internal parameter $u$ and the $c/a$ ratio -- in near-perfect agreement with our DFT results.

We demonstrate that the developed VFF potential accurately captures structural and electronic properties of (B,Ga)N and (B,In,Ga)N alloys across experimentally relevant BN concentrations (2–7\%), with results in excellent agreement with full DFT calculations. Despite not being trained on alloy data, the potential reliably predicts bond lengths, lattice constants, and band gaps.

Beyond enabling large-scale electronic structure simulations of WZ-based alloys and heterostructures, the potential can serve as an efficient pre-relaxation tool for DFT geometry optimizations in large supercells. Its compatibility with the Tanner \emph{et al.}~\cite{TaCa2019} analytic model, without introducing additional force constants, further supports its potential for mixed-phase alloy studies. Thus, our work establishes a transferable framework for constructing VFF models across crystallographic phases.

\begin{acknowledgments}
This work has received funding from  Taighde Eireann—Research Ireland, formerly Science Foundation Ireland
(Grants No. 21/FFP-A/9014, and No. 12/RC/2276 P2). The authors would like to thank Dr. C.~A. Broderick and Dr. C. Murphy for fruitful discussion on the VFF model by Tanner \emph{et al.}~\cite{TaCa2019}.  
\end{acknowledgments}

\section*{Data availability}
The data that support the findings of this article will be made openly available on XXX.

\appendix
\section{Analytic expressions for force constants in ionic ZB crystals}
\label{sec:appendix}

Tanner~\emph{et al.}~\cite{TaCa2019} developed a VFF model based on the approach of Musgrave and Pople~\cite{MuPo1962}, later modified by Martin~\cite{Mart1970}. 
Their formulation explicitly accounts for the Kleinman parameter, $\zeta$, and the three cubic second‑order elastic constants, $C_{11}$, $C_{12}$, and $C_{44}$. They proposed different parametrizations of the potential: a covalent (non‑Coulombic) version for nonionic or weakly ionic materials, and further variants for highly ionic materials. Given that we target III-N materials, we summarize below the analytic expressions of the force constants, expressed in terms of measurable macroscopic elastic constants and inner elastic parameter, $E_{11}$, for highly ionic ZB systems. As outlined in the main text, we use these expressions, along with similarities between the elastic tensors for WZ and [111]-ZB crystals, in our VFF model parametrization. However, it is important to note that our procedure is general and the non-ionic model of Tanner~\emph{et al.}~\cite{TaCa2019} is also compatible with our workflow given in Sec.~\ref{sec:theoretical_framework}, only that the $S$ parameter will not feature.

The analytic expressions for the force constants relevant for our study, see Eqs.~(\ref{eqn:MartinPot}) and~(\ref{eqn:WZPot}), read as follows:

\begin{widetext}
\begin{equation}
    \label{eqn:fc_ktheta}
    k_{\theta} = \frac{2(C_{11} - C_{12} + \frac{3 \sqrt{3}}{16} (\alpha_2 - 2 \alpha_1) S C_{0}) r_0}{3 \sqrt{3}}\,\, ,
\end{equation}

\begin{equation}
    \label{eqn:fc_kr}
    k_{r} = \frac{r_0 [C_{11} (2 + 2\zeta + 5\zeta^{2}) + C_{12} (1 - 8\zeta - 2\zeta^2) + 3C_{44}(1 - 4\zeta) + SC_{0}(a_{1} + a_{2}\zeta + a_{3}\zeta^2 )]}{\sqrt{3} (1 - \zeta)^{2}}\,\, ,
\end{equation}

\begin{equation}
    \label{eqn:fc_krr}
    k_{rr} = \frac{r_0 [C_{11} (2 - 10\zeta - \zeta^{2}) + C_{12} (7 - 8\zeta + 10\zeta^2) - 3C_{44}(1 - 4\zeta) + SC_{0}(a_{4} + a_{5}\zeta + a_{6}\zeta^2 )]}{6\sqrt{3} (1 - \zeta)^{2}}\,\, ,
\end{equation}

\begin{equation}
    \label{eqn:fc_krtheta}
    k_{r\theta} = \frac{r_{0}}{3} \sqrt{\frac{2}{3}} \frac{(C_{11} - C_{12}) (1 + 2\zeta) -3C_{44} +SC_{0} (a_7 + a_8 \zeta)}{\zeta - 1}\,\, ,
\end{equation}
\end{widetext}
where $C_{0}$, in units of GPa, is given by $\frac{e^2}{4 \pi \epsilon_{0} r_{0}^{4}}$. The coefficients $\alpha_{i}$ and the Ewald Summation terms, $a_{i}$, can be found in Ref.~\cite{TaCa2019}; the dimensionless effective charge parameter, $S$, that must be considered for highly ionic materials, is obtained from:
\begin{equation}
    \label{eqn:Sparam}
    S = \frac{E_{11} (1 - \zeta)^2 a_0^2 - 16 (C_{11} - C_{12} - C_{44})}{6 \sqrt{3} C_0 (-\alpha_1 + \alpha_2 + \frac{\alpha_3}{16} - \frac{\alpha_4}{4})}\,\, ,
\end{equation}
where $E_{11}$ describes the internal strain contribution to the free energy. For InN, GaN and AlN we use the same $E_{11}$ value as Ref.~\cite{TaCa2019}. Though not given explicitly in this reference, we can infer the $E_{11}$ value from rearranging Eq.~(\ref{eqn:Sparam}), or directly using Eq. (28) in Ref~\cite{TaCa2019}, as all components are then known for ZB InN, GaN, AlN. For ZB BN these are not known. However, $E_{11}$ can be related to the zone-center transverse optical phonon frequency,   
\begin{equation}
    \label{eqn:E11}
    E_{11} = \frac{4 \mu \omega_\text{TO}^{2}}{a_{0}^3}\,\, .
\end{equation}
Here $\mu$ is the anion-cation reduced mass and $\omega_\text{TO}$ is the transverse optical phonon frequency at the $\Gamma$-point. Thus, the ZB BN value can be calculated from the above and we use $a_0$ and an average of the two $\omega_\text{TO}$ values given in Ref.~\cite{BeGr2000}.

Also, Eq.~(\ref{eqn:fc_ktheta}) corrects a typo in the expression given for $k_\theta$ in Ref.~\cite{TaCa2019}. 

\section{Additional Random Alloy Configurations for B$_{3}$Ga$_{51}$N$_{54}$}
\label{sec:appendixB}

\begin{figure}[H]
    \centering
    \includegraphics{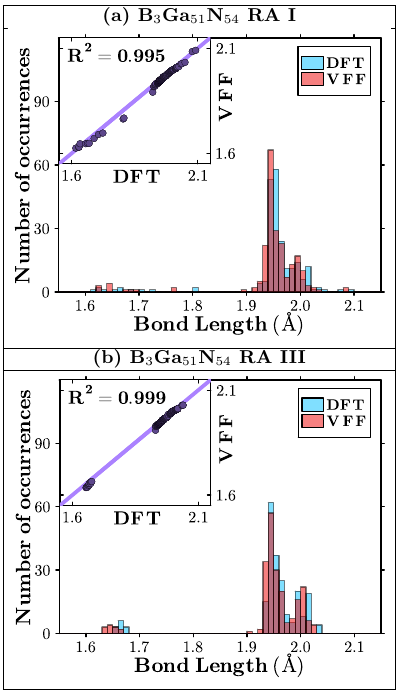}
    \caption{Comparison of bond length distributions from density functional theory (DFT) and valence force field model (VFF) for the the additional B$_{3}$Ga$_{51}$N$_{54}$ random alloy configurations (a) RA I and (b) RA III. As in the main text the histogram bin value is 0.01\AA. Inset: quantile-quantile plots comparing the distributions, where the solid line indicates perfect agreement between VFF and DFT.}
    \label{fig:RAI_RA_II}
\end{figure}

In this appendix, we present details of the bond length distributions of the random alloy configurations RA I and III discussed in the main text for B$_{3}$Ga$_{51}$N$_{54}$. Following the main text, Fig.~\ref{fig:RAI_RA_II} displays the data obtained from VFF and DFT calculuations for (a) RA I and (b) RA III. The insets show Q–Q plots, offering an additional comparison of bond length distributions from VFF and DFT. 

The figure demonstrates that, for both random alloy configurations, the bond‑length distributions predicted by the VFF potential closely match the DFT results. This is further supported by the very high $R^2$ values of $>0.99$, which are even higher than for RA II discussed in Sec.~\ref{subsec:BGaN}. The higher $R^2$ values correlate well with the perfect agreement in band gap values reported in Table.~\ref{tab:alloy_outputs} for RA I and RA III. This demonstrates further the  robustness of our VFF model for performing accurate and reliable structural relaxations in highly ionic and mismatched alloys.

\bibliography{Refs}

\begin{thebibliography}{86}%
\makeatletter
\providecommand \@ifxundefined [1]{%
 \@ifx{#1\undefined}
}%
\providecommand \@ifnum [1]{%
 \ifnum #1\expandafter \@firstoftwo
 \else \expandafter \@secondoftwo
 \fi
}%
\providecommand \@ifx [1]{%
 \ifx #1\expandafter \@firstoftwo
 \else \expandafter \@secondoftwo
 \fi
}%
\providecommand \natexlab [1]{#1}%
\providecommand \enquote  [1]{``#1''}%
\providecommand \bibnamefont  [1]{#1}%
\providecommand \bibfnamefont [1]{#1}%
\providecommand \citenamefont [1]{#1}%
\providecommand \href@noop [0]{\@secondoftwo}%
\providecommand \href [0]{\begingroup \@sanitize@url \@href}%
\providecommand \@href[1]{\@@startlink{#1}\@@href}%
\providecommand \@@href[1]{\endgroup#1\@@endlink}%
\providecommand \@sanitize@url [0]{\catcode `\\12\catcode `\$12\catcode
  `\&12\catcode `\#12\catcode `\^12\catcode `\_12\catcode `\%12\relax}%
\providecommand \@@startlink[1]{}%
\providecommand \@@endlink[0]{}%
\providecommand \url  [0]{\begingroup\@sanitize@url \@url }%
\providecommand \@url [1]{\endgroup\@href {#1}{\urlprefix }}%
\providecommand \urlprefix  [0]{URL }%
\providecommand \Eprint [0]{\href }%
\providecommand \doibase [0]{https://doi.org/}%
\providecommand \selectlanguage [0]{\@gobble}%
\providecommand \bibinfo  [0]{\@secondoftwo}%
\providecommand \bibfield  [0]{\@secondoftwo}%
\providecommand \translation [1]{[#1]}%
\providecommand \BibitemOpen [0]{}%
\providecommand \bibitemStop [0]{}%
\providecommand \bibitemNoStop [0]{.\EOS\space}%
\providecommand \EOS [0]{\spacefactor3000\relax}%
\providecommand \BibitemShut  [1]{\csname bibitem#1\endcsname}%
\let\auto@bib@innerbib\@empty
\bibitem [{\citenamefont {Tanner}\ \emph {et~al.}(2019)\citenamefont {Tanner},
  \citenamefont {Caro}, \citenamefont {Schulz},\ and\ \citenamefont
  {O'Reilly}}]{TaCa2019}%
  \BibitemOpen
  \bibfield  {author} {\bibinfo {author} {\bibfnamefont {D.~S.~P.}\
  \bibnamefont {Tanner}}, \bibinfo {author} {\bibfnamefont {M.~A.}\
  \bibnamefont {Caro}}, \bibinfo {author} {\bibfnamefont {S.}~\bibnamefont
  {Schulz}},\ and\ \bibinfo {author} {\bibfnamefont {E.~P.}\ \bibnamefont
  {O'Reilly}},\ }\bibfield  {title} {\bibinfo {title} {Fully analytic valence
  force field model for the elastic and inner elastic properties of diamond and
  zincblende crystals},\ }\href {https://doi.org/10.1103/PhysRevB.100.094112}
  {\bibfield  {journal} {\bibinfo  {journal} {Phys. Rev. B}\ }\textbf {\bibinfo
  {volume} {100}},\ \bibinfo {pages} {094112} (\bibinfo {year}
  {2019})}\BibitemShut {NoStop}%
\bibitem [{\citenamefont {Simon}()}]{Simo2017}%
  \BibitemOpen
  \bibfield  {author} {\bibinfo {author} {\bibfnamefont {S.~H.}\ \bibnamefont
  {Simon}},\ }\href@noop {} {\emph {\bibinfo {title} {The Oxford solid state
  basics}}},\ \bibinfo {edition} {1st}\ ed.\ (\bibinfo  {publisher} {Oxford
  University Press})\ \bibinfo {note} {{OCLC}: 1091723162}\BibitemShut
  {NoStop}%
\bibitem [{\citenamefont {Banin}\ \emph {et~al.}()\citenamefont {Banin},
  \citenamefont {Cao}, \citenamefont {Katz},\ and\ \citenamefont
  {Millo}}]{BaCa1999}%
  \BibitemOpen
  \bibfield  {author} {\bibinfo {author} {\bibfnamefont {U.}~\bibnamefont
  {Banin}}, \bibinfo {author} {\bibfnamefont {Y.}~\bibnamefont {Cao}}, \bibinfo
  {author} {\bibfnamefont {D.}~\bibnamefont {Katz}},\ and\ \bibinfo {author}
  {\bibfnamefont {O.}~\bibnamefont {Millo}},\ }\bibfield  {title} {\bibinfo
  {title} {Identification of atomic-like electronic states in indium arsenide
  nanocrystal quantum dots},\ }\href {https://doi.org/10.1038/22979} {\bibfield
   {journal} {\bibinfo  {journal} {Nature}\ }\textbf {\bibinfo {volume}
  {400}},\ \bibinfo {pages} {542}}\BibitemShut {NoStop}%
\bibitem [{\citenamefont {Fox}\ and\ \citenamefont
  {Ispasoiu}(2007)}]{FoIs2007}%
  \BibitemOpen
  \bibfield  {author} {\bibinfo {author} {\bibfnamefont {M.}~\bibnamefont
  {Fox}}\ and\ \bibinfo {author} {\bibfnamefont {R.}~\bibnamefont {Ispasoiu}},\
  }\bibinfo {title} {Quantum wells, superlattices, and band-gap engineering},\
  in\ \href {https://doi.org/10.1007/978-0-387-29185-7_42} {\emph {\bibinfo
  {booktitle} {Springer Handbook of Electronic and Photonic Materials}}},\
  \bibinfo {editor} {edited by\ \bibinfo {editor} {\bibfnamefont
  {S.}~\bibnamefont {Kasap}}\ and\ \bibinfo {editor} {\bibfnamefont
  {P.}~\bibnamefont {Capper}}}\ (\bibinfo  {publisher} {Springer US},\ \bibinfo
  {address} {Boston, MA},\ \bibinfo {year} {2007})\ pp.\ \bibinfo {pages}
  {1021--1040}\BibitemShut {NoStop}%
\bibitem [{\citenamefont {O'Donnell}(2005)}]{ODonnell_SemiconductorMaterials}%
  \BibitemOpen
  \bibfield  {author} {\bibinfo {author} {\bibfnamefont {K.}~\bibnamefont
  {O'Donnell}},\ }\bibfield  {title} {\bibinfo {title} {Semiconductor materials
  | iii-nitrides},\ }in\ \href
  {https://doi.org/https://doi.org/10.1016/B0-12-369395-0/00643-6} {\emph
  {\bibinfo {booktitle} {Encyclopedia of Modern Optics}}},\ \bibinfo {editor}
  {edited by\ \bibinfo {editor} {\bibfnamefont {R.~D.}\ \bibnamefont
  {Guenther}}}\ (\bibinfo  {publisher} {Elsevier},\ \bibinfo {address}
  {Oxford},\ \bibinfo {year} {2005})\ pp.\ \bibinfo {pages}
  {372--377}\BibitemShut {NoStop}%
\bibitem [{\citenamefont {Upadhyay}\ and\ \citenamefont
  {Chattopadhyay}(2021)}]{Upadhyay_AlGaN_HEMTSensors}%
  \BibitemOpen
  \bibfield  {author} {\bibinfo {author} {\bibfnamefont {K.~T.}\ \bibnamefont
  {Upadhyay}}\ and\ \bibinfo {author} {\bibfnamefont {M.~K.}\ \bibnamefont
  {Chattopadhyay}},\ }\bibfield  {title} {\bibinfo {title} {Sensor applications
  based on algan/gan heterostructures},\ }\href
  {https://doi.org/https://doi.org/10.1016/j.mseb.2020.114849} {\bibfield
  {journal} {\bibinfo  {journal} {Materials Science and Engineering: B}\
  }\textbf {\bibinfo {volume} {263}},\ \bibinfo {pages} {114849} (\bibinfo
  {year} {2021})}\BibitemShut {NoStop}%
\bibitem [{\citenamefont {Band}\ and\ \citenamefont
  {Avishai}(2013)}]{Low_Dim_QuantumSys}%
  \BibitemOpen
  \bibfield  {author} {\bibinfo {author} {\bibfnamefont {Y.~B.}\ \bibnamefont
  {Band}}\ and\ \bibinfo {author} {\bibfnamefont {Y.}~\bibnamefont {Avishai}},\
  }\bibfield  {title} {\bibinfo {title} {13 - low-dimensional quantum
  systems},\ }in\ \href
  {https://doi.org/https://doi.org/10.1016/B978-0-444-53786-7.00013-7} {\emph
  {\bibinfo {booktitle} {Quantum Mechanics with Applications to Nanotechnology
  and Information Science}}},\ \bibinfo {editor} {edited by\ \bibinfo {editor}
  {\bibfnamefont {Y.~B.}\ \bibnamefont {Band}}\ and\ \bibinfo {editor}
  {\bibfnamefont {Y.}~\bibnamefont {Avishai}}}\ (\bibinfo  {publisher}
  {Academic Press},\ \bibinfo {address} {Amsterdam},\ \bibinfo {year} {2013})\
  pp.\ \bibinfo {pages} {749--823}\BibitemShut {NoStop}%
\bibitem [{\citenamefont {Tidrow}(1997)}]{Tidrow_QWIPs}%
  \BibitemOpen
  \bibfield  {author} {\bibinfo {author} {\bibfnamefont {M.~Z.}\ \bibnamefont
  {Tidrow}},\ }\bibfield  {title} {\bibinfo {title} {Three-well one- and
  two-color quantum well infrared photodetectors},\ }\href
  {https://doi.org/https://doi.org/10.1016/S0254-0584(97)01916-0} {\bibfield
  {journal} {\bibinfo  {journal} {Materials Chemistry and Physics}\ }\textbf
  {\bibinfo {volume} {50}},\ \bibinfo {pages} {183} (\bibinfo {year}
  {1997})}\BibitemShut {NoStop}%
\bibitem [{\citenamefont {Amano}\ \emph {et~al.}(2020)\citenamefont {Amano},
  \citenamefont {Collazo}, \citenamefont {Santi}, \citenamefont {Einfeldt},
  \citenamefont {Funato}, \citenamefont {Glaab}, \citenamefont {Hagedorn},
  \citenamefont {Hirano}, \citenamefont {Hirayama}, \citenamefont {Ishii},
  \citenamefont {Kashima}, \citenamefont {Kawakami}, \citenamefont {Kirste},
  \citenamefont {Kneissl}, \citenamefont {Martin}, \citenamefont {Mehnke},
  \citenamefont {Meneghini}, \citenamefont {Ougazzaden}, \citenamefont
  {Parbrook}, \citenamefont {Rajan}, \citenamefont {Reddy}, \citenamefont
  {Römer}, \citenamefont {Ruschel}, \citenamefont {Sarkar}, \citenamefont
  {Scholz}, \citenamefont {Schowalter}, \citenamefont {Shields}, \citenamefont
  {Sitar}, \citenamefont {Sulmoni}, \citenamefont {Wang}, \citenamefont
  {Wernicke}, \citenamefont {Weyers}, \citenamefont {Witzigmann}, \citenamefont
  {Wu}, \citenamefont {Wunderer},\ and\ \citenamefont {Zhang}}]{UVRoadMap2020}%
  \BibitemOpen
  \bibfield  {author} {\bibinfo {author} {\bibfnamefont {H.}~\bibnamefont
  {Amano}}, \bibinfo {author} {\bibfnamefont {R.}~\bibnamefont {Collazo}},
  \bibinfo {author} {\bibfnamefont {C.~D.}\ \bibnamefont {Santi}}, \bibinfo
  {author} {\bibfnamefont {S.}~\bibnamefont {Einfeldt}}, \bibinfo {author}
  {\bibfnamefont {M.}~\bibnamefont {Funato}}, \bibinfo {author} {\bibfnamefont
  {J.}~\bibnamefont {Glaab}}, \bibinfo {author} {\bibfnamefont
  {S.}~\bibnamefont {Hagedorn}}, \bibinfo {author} {\bibfnamefont
  {A.}~\bibnamefont {Hirano}}, \bibinfo {author} {\bibfnamefont
  {H.}~\bibnamefont {Hirayama}}, \bibinfo {author} {\bibfnamefont
  {R.}~\bibnamefont {Ishii}}, \bibinfo {author} {\bibfnamefont
  {Y.}~\bibnamefont {Kashima}}, \bibinfo {author} {\bibfnamefont
  {Y.}~\bibnamefont {Kawakami}}, \bibinfo {author} {\bibfnamefont
  {R.}~\bibnamefont {Kirste}}, \bibinfo {author} {\bibfnamefont
  {M.}~\bibnamefont {Kneissl}}, \bibinfo {author} {\bibfnamefont
  {R.}~\bibnamefont {Martin}}, \bibinfo {author} {\bibfnamefont
  {F.}~\bibnamefont {Mehnke}}, \bibinfo {author} {\bibfnamefont
  {M.}~\bibnamefont {Meneghini}}, \bibinfo {author} {\bibfnamefont
  {A.}~\bibnamefont {Ougazzaden}}, \bibinfo {author} {\bibfnamefont {P.~J.}\
  \bibnamefont {Parbrook}}, \bibinfo {author} {\bibfnamefont {S.}~\bibnamefont
  {Rajan}}, \bibinfo {author} {\bibfnamefont {P.}~\bibnamefont {Reddy}},
  \bibinfo {author} {\bibfnamefont {F.}~\bibnamefont {Römer}}, \bibinfo
  {author} {\bibfnamefont {J.}~\bibnamefont {Ruschel}}, \bibinfo {author}
  {\bibfnamefont {B.}~\bibnamefont {Sarkar}}, \bibinfo {author} {\bibfnamefont
  {F.}~\bibnamefont {Scholz}}, \bibinfo {author} {\bibfnamefont {L.~J.}\
  \bibnamefont {Schowalter}}, \bibinfo {author} {\bibfnamefont
  {P.}~\bibnamefont {Shields}}, \bibinfo {author} {\bibfnamefont
  {Z.}~\bibnamefont {Sitar}}, \bibinfo {author} {\bibfnamefont
  {L.}~\bibnamefont {Sulmoni}}, \bibinfo {author} {\bibfnamefont
  {T.}~\bibnamefont {Wang}}, \bibinfo {author} {\bibfnamefont {T.}~\bibnamefont
  {Wernicke}}, \bibinfo {author} {\bibfnamefont {M.}~\bibnamefont {Weyers}},
  \bibinfo {author} {\bibfnamefont {B.}~\bibnamefont {Witzigmann}}, \bibinfo
  {author} {\bibfnamefont {Y.~R.}\ \bibnamefont {Wu}}, \bibinfo {author}
  {\bibfnamefont {T.}~\bibnamefont {Wunderer}},\ and\ \bibinfo {author}
  {\bibfnamefont {Y.}~\bibnamefont {Zhang}},\ }\bibfield  {title} {\bibinfo
  {title} {The 2020 uv emitter roadmap},\ }\bibfield  {journal} {\bibinfo
  {journal} {Journal of Physics D: Applied Physics}\ }\textbf {\bibinfo
  {volume} {53}},\ \href {https://doi.org/10.1088/1361-6463/aba64c}
  {10.1088/1361-6463/aba64c} (\bibinfo {year} {2020})\BibitemShut {NoStop}%
\bibitem [{\citenamefont {Mittelstädt}\ \emph {et~al.}()\citenamefont
  {Mittelstädt}, \citenamefont {Schliwa},\ and\ \citenamefont
  {Klenovský}}]{MiSc2022}%
  \BibitemOpen
  \bibfield  {author} {\bibinfo {author} {\bibfnamefont {A.}~\bibnamefont
  {Mittelstädt}}, \bibinfo {author} {\bibfnamefont {A.}~\bibnamefont
  {Schliwa}},\ and\ \bibinfo {author} {\bibfnamefont {P.}~\bibnamefont
  {Klenovský}},\ }\bibfield  {title} {\bibinfo {title} {Modeling electronic
  and optical properties of {III}–v quantum dots—selected recent
  developments},\ }\href {https://doi.org/10.1038/s41377-021-00700-9}
  {\bibfield  {journal} {\bibinfo  {journal} {Light: Science \& Applications}\
  }\textbf {\bibinfo {volume} {11}},\ \bibinfo {pages} {17}}\BibitemShut
  {NoStop}%
\bibitem [{\citenamefont {Kumar}\ \emph {et~al.}(2025)\citenamefont {Kumar},
  \citenamefont {Rai},\ and\ \citenamefont {Askari}}]{KuRa2025}%
  \BibitemOpen
  \bibfield  {author} {\bibinfo {author} {\bibfnamefont {S.}~\bibnamefont
  {Kumar}}, \bibinfo {author} {\bibfnamefont {A.}~\bibnamefont {Rai}},\ and\
  \bibinfo {author} {\bibfnamefont {S.~S.~A.}\ \bibnamefont {Askari}},\
  }\bibfield  {title} {\bibinfo {title} {Interdiffusion induced changes in the
  absorption spectra of iii-v quantum dot systems},\ }\href
  {https://doi.org/https://doi.org/10.1016/j.ijleo.2024.172159} {\bibfield
  {journal} {\bibinfo  {journal} {Optik}\ }\textbf {\bibinfo {volume} {321}},\
  \bibinfo {pages} {172159} (\bibinfo {year} {2025})}\BibitemShut {NoStop}%
\bibitem [{\citenamefont {Schulz}\ \emph {et~al.}(2015)\citenamefont {Schulz},
  \citenamefont {Caro}, \citenamefont {Coughlan},\ and\ \citenamefont
  {O'Reilly}}]{ScCa2015}%
  \BibitemOpen
  \bibfield  {author} {\bibinfo {author} {\bibfnamefont {S.}~\bibnamefont
  {Schulz}}, \bibinfo {author} {\bibfnamefont {M.~A.}\ \bibnamefont {Caro}},
  \bibinfo {author} {\bibfnamefont {C.}~\bibnamefont {Coughlan}},\ and\
  \bibinfo {author} {\bibfnamefont {E.~P.}\ \bibnamefont {O'Reilly}},\
  }\bibfield  {title} {\bibinfo {title} {Atomistic analysis of the impact of
  alloy and well-width fluctuations on the electronic and optical properties of
  ingan/gan quantum wells},\ }\href
  {https://doi.org/10.1103/PhysRevB.91.035439} {\bibfield  {journal} {\bibinfo
  {journal} {Phys. Rev. B}\ }\textbf {\bibinfo {volume} {91}},\ \bibinfo
  {pages} {035439} (\bibinfo {year} {2015})}\BibitemShut {NoStop}%
\bibitem [{\citenamefont {Walukiewicz}\ and\ \citenamefont
  {Zide}()}]{WaZi2020}%
  \BibitemOpen
  \bibfield  {author} {\bibinfo {author} {\bibfnamefont {W.}~\bibnamefont
  {Walukiewicz}}\ and\ \bibinfo {author} {\bibfnamefont {J.~M.~O.}\
  \bibnamefont {Zide}},\ }\bibfield  {title} {\bibinfo {title} {Highly
  mismatched semiconductor alloys: From atoms to devices},\ }\href
  {https://doi.org/10.1063/1.5142248} {\bibfield  {journal} {\bibinfo
  {journal} {Journal of Applied Physics}\ }\textbf {\bibinfo {volume} {127}},\
  \bibinfo {pages} {010401}}\BibitemShut {NoStop}%
\bibitem [{\citenamefont {Manzeli}\ \emph {et~al.}()\citenamefont {Manzeli},
  \citenamefont {Ovchinnikov}, \citenamefont {Pasquier}, \citenamefont
  {Yazyev},\ and\ \citenamefont {Kis}}]{MaDm2017}%
  \BibitemOpen
  \bibfield  {author} {\bibinfo {author} {\bibfnamefont {S.}~\bibnamefont
  {Manzeli}}, \bibinfo {author} {\bibfnamefont {D.}~\bibnamefont
  {Ovchinnikov}}, \bibinfo {author} {\bibfnamefont {D.}~\bibnamefont
  {Pasquier}}, \bibinfo {author} {\bibfnamefont {O.~V.}\ \bibnamefont
  {Yazyev}},\ and\ \bibinfo {author} {\bibfnamefont {A.}~\bibnamefont {Kis}},\
  }\bibfield  {title} {\bibinfo {title} {2d transition metal dichalcogenides},\
  }\href {https://doi.org/10.1038/natrevmats.2017.33} {\bibfield  {journal}
  {\bibinfo  {journal} {Nature Reviews Materials}\ }\textbf {\bibinfo {volume}
  {2}},\ \bibinfo {pages} {17033}}\BibitemShut {NoStop}%
\bibitem [{\citenamefont {Chen}\ \emph {et~al.}()\citenamefont {Chen},
  \citenamefont {Li}, \citenamefont {Li},\ and\ \citenamefont {Li}}]{ChLi2020}%
  \BibitemOpen
  \bibfield  {author} {\bibinfo {author} {\bibfnamefont {W.}~\bibnamefont
  {Chen}}, \bibinfo {author} {\bibfnamefont {X.}~\bibnamefont {Li}}, \bibinfo
  {author} {\bibfnamefont {Y.}~\bibnamefont {Li}},\ and\ \bibinfo {author}
  {\bibfnamefont {Y.}~\bibnamefont {Li}},\ }\bibfield  {title} {\bibinfo
  {title} {A review: crystal growth for high-performance all-inorganic
  perovskite solar cells},\ }\href {https://doi.org/10.1039/D0EE00215A}
  {\bibfield  {journal} {\bibinfo  {journal} {Energy \& Environmental Science}\
  }\textbf {\bibinfo {volume} {13}},\ \bibinfo {pages} {1971}}\BibitemShut
  {NoStop}%
\bibitem [{\citenamefont {Lopez-Varo}\ \emph {et~al.}()\citenamefont
  {Lopez-Varo}, \citenamefont {Jiménez-Tejada}, \citenamefont
  {García-Rosell}, \citenamefont {Ravishankar}, \citenamefont
  {Garcia-Belmonte}, \citenamefont {Bisquert},\ and\ \citenamefont
  {Almora}}]{LoJi2018}%
  \BibitemOpen
  \bibfield  {author} {\bibinfo {author} {\bibfnamefont {P.}~\bibnamefont
  {Lopez-Varo}}, \bibinfo {author} {\bibfnamefont {J.~A.}\ \bibnamefont
  {Jiménez-Tejada}}, \bibinfo {author} {\bibfnamefont {M.}~\bibnamefont
  {García-Rosell}}, \bibinfo {author} {\bibfnamefont {S.}~\bibnamefont
  {Ravishankar}}, \bibinfo {author} {\bibfnamefont {G.}~\bibnamefont
  {Garcia-Belmonte}}, \bibinfo {author} {\bibfnamefont {J.}~\bibnamefont
  {Bisquert}},\ and\ \bibinfo {author} {\bibfnamefont {O.}~\bibnamefont
  {Almora}},\ }\bibfield  {title} {\bibinfo {title} {Device physics of hybrid
  perovskite solar cells: Theory and experiment},\ }\href
  {https://doi.org/10.1002/aenm.201702772} {\bibfield  {journal} {\bibinfo
  {journal} {Advanced Energy Materials}\ }\textbf {\bibinfo {volume} {8}},\
  \bibinfo {pages} {1702772}}\BibitemShut {NoStop}%
\bibitem [{\citenamefont {Suckert}\ \emph {et~al.}(2021)\citenamefont
  {Suckert}, \citenamefont {R{\"o}dl}, \citenamefont {Furthm{\"u}ller},
  \citenamefont {Bechstedt},\ and\ \citenamefont {Botti}}]{SuRo2021}%
  \BibitemOpen
  \bibfield  {author} {\bibinfo {author} {\bibfnamefont {J.~R.}\ \bibnamefont
  {Suckert}}, \bibinfo {author} {\bibfnamefont {C.}~\bibnamefont {R{\"o}dl}},
  \bibinfo {author} {\bibfnamefont {J.}~\bibnamefont {Furthm{\"u}ller}},
  \bibinfo {author} {\bibfnamefont {F.}~\bibnamefont {Bechstedt}},\ and\
  \bibinfo {author} {\bibfnamefont {S.}~\bibnamefont {Botti}},\ }\bibfield
  {title} {\bibinfo {title} {Efficient strain-induced light emission in
  lonsdaleite germanium},\ }\href
  {https://doi.org/10.1103/PhysRevMaterials.5.024602} {\bibfield  {journal}
  {\bibinfo  {journal} {Phys. Rev. Materials}\ }\textbf {\bibinfo {volume}
  {5}},\ \bibinfo {pages} {024602} (\bibinfo {year} {2021})}\BibitemShut
  {NoStop}%
\bibitem [{\citenamefont {De}\ and\ \citenamefont {Pryor}()}]{DePr2014}%
  \BibitemOpen
  \bibfield  {author} {\bibinfo {author} {\bibfnamefont {A.}~\bibnamefont
  {De}}\ and\ \bibinfo {author} {\bibfnamefont {C.~E.}\ \bibnamefont {Pryor}},\
  }\bibfield  {title} {\bibinfo {title} {Electronic structure and optical
  properties of si, ge and diamond in the lonsdaleite phase},\ }\href
  {https://doi.org/10.1088/0953-8984/26/4/045801} {\bibfield  {journal}
  {\bibinfo  {journal} {Journal of Physics: Condensed Matter}\ }\textbf
  {\bibinfo {volume} {26}},\ \bibinfo {pages} {045801}}\BibitemShut {NoStop}%
\bibitem [{\citenamefont {Peeters}\ \emph {et~al.}(2024)\citenamefont
  {Peeters}, \citenamefont {{van Lange}}, \citenamefont {Belabbes},
  \citenamefont {{van Hemert}}, \citenamefont {Jansen}, \citenamefont {Farina},
  \citenamefont {{van Tilburg}}, \citenamefont {Verheijen}, \citenamefont
  {Botti}, \citenamefont {Bechstedt}, \citenamefont {Haverkort},\ and\
  \citenamefont {Bakkers}}]{PevLa2024}%
  \BibitemOpen
  \bibfield  {author} {\bibinfo {author} {\bibfnamefont {W.~H.~J.}\
  \bibnamefont {Peeters}}, \bibinfo {author} {\bibfnamefont {V.~T.}\
  \bibnamefont {{van Lange}}}, \bibinfo {author} {\bibfnamefont
  {A.}~\bibnamefont {Belabbes}}, \bibinfo {author} {\bibfnamefont {M.~C.}\
  \bibnamefont {{van Hemert}}}, \bibinfo {author} {\bibfnamefont {M.~M.}\
  \bibnamefont {Jansen}}, \bibinfo {author} {\bibfnamefont {R.}~\bibnamefont
  {Farina}}, \bibinfo {author} {\bibfnamefont {M.~A.~J.}\ \bibnamefont {{van
  Tilburg}}}, \bibinfo {author} {\bibfnamefont {M.~A.}\ \bibnamefont
  {Verheijen}}, \bibinfo {author} {\bibfnamefont {S.}~\bibnamefont {Botti}},
  \bibinfo {author} {\bibfnamefont {F.}~\bibnamefont {Bechstedt}}, \bibinfo
  {author} {\bibfnamefont {J.~E.~M.}\ \bibnamefont {Haverkort}},\ and\ \bibinfo
  {author} {\bibfnamefont {E.~P.~A.~M.}\ \bibnamefont {Bakkers}},\ }\bibfield
  {title} {\bibinfo {title} {Direct bandgap quantum wells in hexagonal silicon
  germanium},\ }\href {https://doi.org/10.1038/s41467-024-49399-3} {\bibfield
  {journal} {\bibinfo  {journal} {Nat. Commun.}\ }\textbf {\bibinfo {volume}
  {15}},\ \bibinfo {pages} {5252} (\bibinfo {year} {2024})}\BibitemShut
  {NoStop}%
\bibitem [{\citenamefont {Leandro}\ \emph {et~al.}(2020)\citenamefont
  {Leandro}, \citenamefont {Reznik}, \citenamefont {Clement}, \citenamefont
  {Repän}, \citenamefont {Reynolds}, \citenamefont {Ubyivovk}, \citenamefont
  {Shtrom}, \citenamefont {Cirlin},\ and\ \citenamefont {Akopian}}]{LeRe2020}%
  \BibitemOpen
  \bibfield  {author} {\bibinfo {author} {\bibfnamefont {L.}~\bibnamefont
  {Leandro}}, \bibinfo {author} {\bibfnamefont {R.}~\bibnamefont {Reznik}},
  \bibinfo {author} {\bibfnamefont {J.~D.}\ \bibnamefont {Clement}}, \bibinfo
  {author} {\bibfnamefont {J.}~\bibnamefont {Repän}}, \bibinfo {author}
  {\bibfnamefont {M.}~\bibnamefont {Reynolds}}, \bibinfo {author}
  {\bibfnamefont {E.~V.}\ \bibnamefont {Ubyivovk}}, \bibinfo {author}
  {\bibfnamefont {I.~V.}\ \bibnamefont {Shtrom}}, \bibinfo {author}
  {\bibfnamefont {G.}~\bibnamefont {Cirlin}},\ and\ \bibinfo {author}
  {\bibfnamefont {N.}~\bibnamefont {Akopian}},\ }\bibfield  {title} {\bibinfo
  {title} {Wurtzite {AlGaAs} nanowires},\ }\href
  {https://doi.org/10.1038/s41598-020-57563-0} {\bibfield  {journal} {\bibinfo
  {journal} {Scientific Reports}\ }\textbf {\bibinfo {volume} {10}},\ \bibinfo
  {pages} {735} (\bibinfo {year} {2020})}\BibitemShut {NoStop}%
\bibitem [{\citenamefont {Sink}\ and\ \citenamefont {Pryor}(2023)}]{SiPr2023}%
  \BibitemOpen
  \bibfield  {author} {\bibinfo {author} {\bibfnamefont {J.}~\bibnamefont
  {Sink}}\ and\ \bibinfo {author} {\bibfnamefont {C.}~\bibnamefont {Pryor}},\
  }\bibfield  {title} {\bibinfo {title} {Wurtzite/zinc-blende crystal-phase
  gaas heterostructures in the tight-binding approximation},\ }\href
  {https://doi.org/10.1103/PhysRevB.108.075104} {\bibfield  {journal} {\bibinfo
   {journal} {Phys. Rev. B}\ }\textbf {\bibinfo {volume} {108}},\ \bibinfo
  {pages} {075104} (\bibinfo {year} {2023})}\BibitemShut {NoStop}%
\bibitem [{\citenamefont {Adachi}()}]{Adac1994}%
  \BibitemOpen
  \bibfield  {author} {\bibinfo {author} {\bibfnamefont {S.}~\bibnamefont
  {Adachi}},\ }\href {https://doi.org/10.1142/2508} {\emph {\bibinfo {title}
  {{GaAs} and Related Materials: Bulk Semiconducting and Superlattice
  Properties}}}\ (\bibinfo  {publisher} {World Scientific})\BibitemShut
  {NoStop}%
\bibitem [{\citenamefont {Jiang}\ \emph {et~al.}()\citenamefont {Jiang},
  \citenamefont {Xiao}, \citenamefont {Peng}, \citenamefont {Qiao},
  \citenamefont {Yang}, \citenamefont {Liu},\ and\ \citenamefont
  {Zu}}]{JiXi2018}%
  \BibitemOpen
  \bibfield  {author} {\bibinfo {author} {\bibfnamefont {M.}~\bibnamefont
  {Jiang}}, \bibinfo {author} {\bibfnamefont {H.}~\bibnamefont {Xiao}},
  \bibinfo {author} {\bibfnamefont {S.}~\bibnamefont {Peng}}, \bibinfo {author}
  {\bibfnamefont {L.}~\bibnamefont {Qiao}}, \bibinfo {author} {\bibfnamefont
  {G.}~\bibnamefont {Yang}}, \bibinfo {author} {\bibfnamefont {Z.}~\bibnamefont
  {Liu}},\ and\ \bibinfo {author} {\bibfnamefont {X.}~\bibnamefont {Zu}},\
  }\bibfield  {title} {\bibinfo {title} {First-principles study of point
  defects in {GaAs}/{AlAs} superlattice: the phase stability and the effects on
  the band structure and carrier mobility},\ }\href
  {https://doi.org/10.1186/s11671-018-2719-7} {\bibfield  {journal} {\bibinfo
  {journal} {Nanoscale Research Letters}\ }\textbf {\bibinfo {volume} {13}},\
  \bibinfo {pages} {301}}\BibitemShut {NoStop}%
\bibitem [{\citenamefont {De}\ and\ \citenamefont {Pryor}(2010)}]{DePr2010}%
  \BibitemOpen
  \bibfield  {author} {\bibinfo {author} {\bibfnamefont {A.}~\bibnamefont
  {De}}\ and\ \bibinfo {author} {\bibfnamefont {C.~E.}\ \bibnamefont {Pryor}},\
  }\bibfield  {title} {\bibinfo {title} {Predicted band structures of {III}-v
  semiconductors in the wurtzite phase},\ }\href
  {https://doi.org/10.1103/PhysRevB.81.155210} {\bibfield  {journal} {\bibinfo
  {journal} {Physical Review B}\ }\textbf {\bibinfo {volume} {81}},\ \bibinfo
  {pages} {155210} (\bibinfo {year} {2010})}\BibitemShut {NoStop}%
\bibitem [{\citenamefont {Zhang}\ \emph {et~al.}(2000)\citenamefont {Zhang},
  \citenamefont {Iqbal}, \citenamefont {Vijayalakshmi}, \citenamefont {Qadri},\
  and\ \citenamefont {Grebel}}]{ZhIq2000}%
  \BibitemOpen
  \bibfield  {author} {\bibinfo {author} {\bibfnamefont {Y.}~\bibnamefont
  {Zhang}}, \bibinfo {author} {\bibfnamefont {Z.}~\bibnamefont {Iqbal}},
  \bibinfo {author} {\bibfnamefont {S.}~\bibnamefont {Vijayalakshmi}}, \bibinfo
  {author} {\bibfnamefont {S.}~\bibnamefont {Qadri}},\ and\ \bibinfo {author}
  {\bibfnamefont {H.}~\bibnamefont {Grebel}},\ }\bibfield  {title} {\bibinfo
  {title} {Formation of hexagonal-wurtzite germanium by pulsed laser
  ablation},\ }\href
  {https://doi.org/https://doi.org/10.1016/S0038-1098(00)00259-3} {\bibfield
  {journal} {\bibinfo  {journal} {Solid State Communications}\ }\textbf
  {\bibinfo {volume} {115}},\ \bibinfo {pages} {657} (\bibinfo {year}
  {2000})}\BibitemShut {NoStop}%
\bibitem [{\citenamefont {Bhullar}\ \emph {et~al.}(2024)\citenamefont
  {Bhullar}, \citenamefont {Akinpelu},\ and\ \citenamefont {Yao}}]{BhAk2024}%
  \BibitemOpen
  \bibfield  {author} {\bibinfo {author} {\bibfnamefont {M.}~\bibnamefont
  {Bhullar}}, \bibinfo {author} {\bibfnamefont {A.}~\bibnamefont {Akinpelu}},\
  and\ \bibinfo {author} {\bibfnamefont {Y.}~\bibnamefont {Yao}},\ }\bibfield
  {title} {\bibinfo {title} {Unveiling novel direct bandgap allotropes of
  germanium: A computational exploration},\ }\href
  {https://doi.org/https://doi.org/10.1016/j.commt.2024.100009} {\bibfield
  {journal} {\bibinfo  {journal} {Computational Materials Today}\ }\textbf
  {\bibinfo {volume} {2-3}},\ \bibinfo {pages} {100009} (\bibinfo {year}
  {2024})}\BibitemShut {NoStop}%
\bibitem [{\citenamefont {Soref}(2006)}]{Sore2006}%
  \BibitemOpen
  \bibfield  {author} {\bibinfo {author} {\bibfnamefont {R.}~\bibnamefont
  {Soref}},\ }\bibfield  {title} {\bibinfo {title} {The past, present, and
  future of silicon photonic},\ }\href
  {https://doi.org/10.1109/JSTQE.2006.883151} {\bibfield  {journal} {\bibinfo
  {journal} {IEEE J. Sel. Top. Quantum Electron.}\ }\textbf {\bibinfo {volume}
  {12}},\ \bibinfo {pages} {1678–1687} (\bibinfo {year} {2006})}\BibitemShut
  {NoStop}%
\bibitem [{\citenamefont {Kudrawiec}\ and\ \citenamefont
  {Hommel}(2020)}]{KuRo2020}%
  \BibitemOpen
  \bibfield  {author} {\bibinfo {author} {\bibfnamefont {R.}~\bibnamefont
  {Kudrawiec}}\ and\ \bibinfo {author} {\bibfnamefont {D.}~\bibnamefont
  {Hommel}},\ }\bibfield  {title} {\bibinfo {title} {{Bandgap engineering in
  III-nitrides with boron and group V elements: Toward applications in
  ultraviolet emitters}},\ }\href {https://doi.org/10.1063/5.0025371}
  {\bibfield  {journal} {\bibinfo  {journal} {Applied Physics Reviews}\
  }\textbf {\bibinfo {volume} {7}},\ \bibinfo {pages} {041314} (\bibinfo {year}
  {2020})},\ \Eprint
  {https://arxiv.org/abs/https://pubs.aip.org/aip/apr/article-pdf/doi/10.1063/5.0025371/13896527/041314\_1\_online.pdf}
  {https://pubs.aip.org/aip/apr/article-pdf/doi/10.1063/5.0025371/13896527/041314\_1\_online.pdf}
  \BibitemShut {NoStop}%
\bibitem [{\citenamefont {Gautier}\ \emph {et~al.}(2010)\citenamefont
  {Gautier}, \citenamefont {Orsal}, \citenamefont {Moudakir}, \citenamefont
  {Maloufi}, \citenamefont {Jomard}, \citenamefont {Alnot}, \citenamefont
  {Djebbour}, \citenamefont {Sirenko}, \citenamefont {Abid}, \citenamefont
  {Pantzas}, \citenamefont {Ferguson}, \citenamefont {Voss},\ and\
  \citenamefont {Ougazzaden}}]{GaOr2010}%
  \BibitemOpen
  \bibfield  {author} {\bibinfo {author} {\bibfnamefont {S.}~\bibnamefont
  {Gautier}}, \bibinfo {author} {\bibfnamefont {G.}~\bibnamefont {Orsal}},
  \bibinfo {author} {\bibfnamefont {T.}~\bibnamefont {Moudakir}}, \bibinfo
  {author} {\bibfnamefont {N.}~\bibnamefont {Maloufi}}, \bibinfo {author}
  {\bibfnamefont {F.}~\bibnamefont {Jomard}}, \bibinfo {author} {\bibfnamefont
  {M.}~\bibnamefont {Alnot}}, \bibinfo {author} {\bibfnamefont
  {Z.}~\bibnamefont {Djebbour}}, \bibinfo {author} {\bibfnamefont
  {A.}~\bibnamefont {Sirenko}}, \bibinfo {author} {\bibfnamefont
  {M.}~\bibnamefont {Abid}}, \bibinfo {author} {\bibfnamefont {K.}~\bibnamefont
  {Pantzas}}, \bibinfo {author} {\bibfnamefont {I.}~\bibnamefont {Ferguson}},
  \bibinfo {author} {\bibfnamefont {P.}~\bibnamefont {Voss}},\ and\ \bibinfo
  {author} {\bibfnamefont {A.}~\bibnamefont {Ougazzaden}},\ }\bibfield  {title}
  {\bibinfo {title} {Metal-organic vapour phase epitaxy of bingan quaternary
  alloys and characterization of boron content},\ }\href
  {https://doi.org/https://doi.org/10.1016/j.jcrysgro.2009.11.040} {\bibfield
  {journal} {\bibinfo  {journal} {Journal of Crystal Growth}\ }\textbf
  {\bibinfo {volume} {312}},\ \bibinfo {pages} {641} (\bibinfo {year}
  {2010})}\BibitemShut {NoStop}%
\bibitem [{\citenamefont {Park}\ and\ \citenamefont {Ahn}(2016)}]{PaAh2016}%
  \BibitemOpen
  \bibfield  {author} {\bibinfo {author} {\bibfnamefont {S.}~\bibnamefont
  {Park}}\ and\ \bibinfo {author} {\bibfnamefont {D.}~\bibnamefont {Ahn}},\
  }\bibfield  {title} {\bibinfo {title} {Effect of boron incorporation on light
  emission characteristics of uv balgan/aln quantum well structures},\ }\href
  {https://doi.org/10.7567/APEX.9.021001} {\bibfield  {journal} {\bibinfo
  {journal} {Applied Physics Express}\ }\textbf {\bibinfo {volume} {9}},\
  \bibinfo {pages} {021001} (\bibinfo {year} {2016})}\BibitemShut {NoStop}%
\bibitem [{\citenamefont {Park}\ and\ \citenamefont {Ahn}(2019)}]{PaAh2019}%
  \BibitemOpen
  \bibfield  {author} {\bibinfo {author} {\bibfnamefont {S.}~\bibnamefont
  {Park}}\ and\ \bibinfo {author} {\bibfnamefont {D.}~\bibnamefont {Ahn}},\
  }\bibfield  {title} {\bibinfo {title} {Theoretical study of optical
  properties of non-polar balgan/aln quantum wells lattice-matched to aln},\
  }\href {https://doi.org/https://doi.org/10.1016/j.ssc.2019.01.006} {\bibfield
   {journal} {\bibinfo  {journal} {Solid State Communications}\ }\textbf
  {\bibinfo {volume} {290}},\ \bibinfo {pages} {67} (\bibinfo {year}
  {2019})}\BibitemShut {NoStop}%
\bibitem [{\citenamefont {Williams}\ and\ \citenamefont
  {Kioupakis}(2019)}]{LoKi2019}%
  \BibitemOpen
  \bibfield  {author} {\bibinfo {author} {\bibfnamefont {L.}~\bibnamefont
  {Williams}}\ and\ \bibinfo {author} {\bibfnamefont {E.}~\bibnamefont
  {Kioupakis}},\ }\bibfield  {title} {\bibinfo {title} {{BAlGaN alloys nearly
  lattice-matched to AlN for efficient UV LEDs}},\ }\href
  {https://doi.org/10.1063/1.5129387} {\bibfield  {journal} {\bibinfo
  {journal} {Applied Physics Letters}\ }\textbf {\bibinfo {volume} {115}},\
  \bibinfo {pages} {231103} (\bibinfo {year} {2019})},\ \Eprint
  {https://arxiv.org/abs/https://pubs.aip.org/aip/apl/article-pdf/doi/10.1063/1.5129387/13314763/231103\_1\_online.pdf}
  {https://pubs.aip.org/aip/apl/article-pdf/doi/10.1063/1.5129387/13314763/231103\_1\_online.pdf}
  \BibitemShut {NoStop}%
\bibitem [{\citenamefont {Solozhenko}\ \emph {et~al.}(1999)\citenamefont
  {Solozhenko}, \citenamefont {Turkevich},\ and\ \citenamefont
  {Holzapfel}}]{SoTu1999}%
  \BibitemOpen
  \bibfield  {author} {\bibinfo {author} {\bibfnamefont {V.~L.}\ \bibnamefont
  {Solozhenko}}, \bibinfo {author} {\bibfnamefont {V.~Z.}\ \bibnamefont
  {Turkevich}},\ and\ \bibinfo {author} {\bibfnamefont {W.~B.}\ \bibnamefont
  {Holzapfel}},\ }\bibfield  {title} {\bibinfo {title} {Refined phase diagram
  of boron nitride},\ }\href {https://doi.org/10.1021/jp984682c} {\bibfield
  {journal} {\bibinfo  {journal} {The Journal of Physical Chemistry B}\
  }\textbf {\bibinfo {volume} {103}},\ \bibinfo {pages} {2903} (\bibinfo {year}
  {1999})}\BibitemShut {NoStop}%
\bibitem [{\citenamefont {Gunning}\ \emph {et~al.}(2017)\citenamefont
  {Gunning}, \citenamefont {Moseley}, \citenamefont {Koleske}, \citenamefont
  {Allerman},\ and\ \citenamefont {Lee}}]{GuMo2017}%
  \BibitemOpen
  \bibfield  {author} {\bibinfo {author} {\bibfnamefont {B.~P.}\ \bibnamefont
  {Gunning}}, \bibinfo {author} {\bibfnamefont {M.~W.}\ \bibnamefont
  {Moseley}}, \bibinfo {author} {\bibfnamefont {D.~D.}\ \bibnamefont
  {Koleske}}, \bibinfo {author} {\bibfnamefont {A.~A.}\ \bibnamefont
  {Allerman}},\ and\ \bibinfo {author} {\bibfnamefont {S.~R.}\ \bibnamefont
  {Lee}},\ }\bibfield  {title} {\bibinfo {title} {Phase degradation in
  bxga1\-xn films grown at low temperature by metalorganic vapor phase
  epitaxy},\ }\href
  {https://doi.org/https://doi.org/10.1016/j.jcrysgro.2016.10.054} {\bibfield
  {journal} {\bibinfo  {journal} {Journal of Crystal Growth}\ }\textbf
  {\bibinfo {volume} {464}},\ \bibinfo {pages} {190} (\bibinfo {year}
  {2017})},\ \bibinfo {note} {proceedings of the 18th International Conference
  on Metal Organic Vapor Phase Epitaxy}\BibitemShut {NoStop}%
\bibitem [{\citenamefont {Lin}\ \emph {et~al.}()\citenamefont {Lin},
  \citenamefont {Liu}, \citenamefont {Liu}, \citenamefont {Lu}, \citenamefont
  {Liu},\ and\ \citenamefont {Li}}]{LiLi2020}%
  \BibitemOpen
  \bibfield  {author} {\bibinfo {author} {\bibfnamefont {R.}~\bibnamefont
  {Lin}}, \bibinfo {author} {\bibfnamefont {X.}~\bibnamefont {Liu}}, \bibinfo
  {author} {\bibfnamefont {K.}~\bibnamefont {Liu}}, \bibinfo {author}
  {\bibfnamefont {Y.}~\bibnamefont {Lu}}, \bibinfo {author} {\bibfnamefont
  {X.}~\bibnamefont {Liu}},\ and\ \bibinfo {author} {\bibfnamefont
  {X.}~\bibnamefont {Li}},\ }\bibfield  {title} {\bibinfo {title} {{BAlN} alloy
  for enhanced two-dimensional electron gas characteristics of {GaN}/{AlGaN}
  heterostructures},\ }\href {https://doi.org/10.1088/1361-6463/aba4d5}
  {\bibfield  {journal} {\bibinfo  {journal} {Journal of Physics D: Applied
  Physics}\ }\textbf {\bibinfo {volume} {53}},\ \bibinfo {pages}
  {48LT01}}\BibitemShut {NoStop}%
\bibitem [{\citenamefont {Tran}\ \emph {et~al.}(2020)\citenamefont {Tran},
  \citenamefont {Liao}, \citenamefont {AlQatari},\ and\ \citenamefont
  {Li}}]{TrLi2020}%
  \BibitemOpen
  \bibfield  {author} {\bibinfo {author} {\bibfnamefont {T.~B.}\ \bibnamefont
  {Tran}}, \bibinfo {author} {\bibfnamefont {C.-H.}\ \bibnamefont {Liao}},
  \bibinfo {author} {\bibfnamefont {F.}~\bibnamefont {AlQatari}},\ and\
  \bibinfo {author} {\bibfnamefont {X.}~\bibnamefont {Li}},\ }\bibfield
  {title} {\bibinfo {title} {Demonstration of single-phase wurtzite baln with
  over 20\% boron content by metalorganic chemical vapor deposition},\ }\href
  {https://doi.org/10.1063/5.0019881} {\bibfield  {journal} {\bibinfo
  {journal} {Applied Physics Letters}\ }\textbf {\bibinfo {volume} {117}},\
  \bibinfo {pages} {082102} (\bibinfo {year} {2020})}\BibitemShut {NoStop}%
\bibitem [{\citenamefont {Sarker}\ and\ \citenamefont
  {Mazumder}(2021)}]{SaMs2021}%
  \BibitemOpen
  \bibfield  {author} {\bibinfo {author} {\bibfnamefont {J.}~\bibnamefont
  {Sarker}}\ and\ \bibinfo {author} {\bibfnamefont {B.}~\bibnamefont
  {Mazumder}},\ }\bibfield  {title} {\bibinfo {title} {A comprehensive review
  on the effects of local microstructures and nanoscale chemical features on
  b-iii-nitride films},\ }\href {https://doi.org/10.1557/s43578-021-00340-0}
  {\bibfield  {journal} {\bibinfo  {journal} {Journal of Materials Research}\
  }\textbf {\bibinfo {volume} {36}},\ \bibinfo {pages} {4665} (\bibinfo {year}
  {2021})}\BibitemShut {NoStop}%
\bibitem [{\citenamefont {Broderick}\ \emph {et~al.}()\citenamefont
  {Broderick}, \citenamefont {O’Reilly},\ and\ \citenamefont
  {Schulz}}]{BrOR2024}%
  \BibitemOpen
  \bibfield  {author} {\bibinfo {author} {\bibfnamefont {C.~A.}\ \bibnamefont
  {Broderick}}, \bibinfo {author} {\bibfnamefont {E.~P.}\ \bibnamefont
  {O’Reilly}},\ and\ \bibinfo {author} {\bibfnamefont {S.}~\bibnamefont
  {Schulz}},\ }\bibfield  {title} {\bibinfo {title} {Perspective: Theory and
  simulation of highly mismatched semiconductor alloys using the tight-binding
  method},\ }\href {https://doi.org/10.1063/5.0192047} {\bibfield  {journal}
  {\bibinfo  {journal} {Journal of Applied Physics}\ }\textbf {\bibinfo
  {volume} {135}},\ \bibinfo {pages} {100902}}\BibitemShut {NoStop}%
\bibitem [{\citenamefont {Jaquez}\ \emph {et~al.}()\citenamefont {Jaquez},
  \citenamefont {Specht}, \citenamefont {Yu}, \citenamefont {Walukiewicz},\
  and\ \citenamefont {Dubon}}]{JaSp2019}%
  \BibitemOpen
  \bibfield  {author} {\bibinfo {author} {\bibfnamefont {M.}~\bibnamefont
  {Jaquez}}, \bibinfo {author} {\bibfnamefont {P.}~\bibnamefont {Specht}},
  \bibinfo {author} {\bibfnamefont {K.~M.}\ \bibnamefont {Yu}}, \bibinfo
  {author} {\bibfnamefont {W.}~\bibnamefont {Walukiewicz}},\ and\ \bibinfo
  {author} {\bibfnamefont {O.~D.}\ \bibnamefont {Dubon}},\ }\bibfield  {title}
  {\bibinfo {title} {Amorphous gallium oxide sulfide: A highly mismatched
  alloy},\ }\href {https://doi.org/10.1063/1.5111985} {\bibfield  {journal}
  {\bibinfo  {journal} {Journal of Applied Physics}\ }\textbf {\bibinfo
  {volume} {126}},\ \bibinfo {pages} {105708}}\BibitemShut {NoStop}%
\bibitem [{\citenamefont {Yu}\ \emph {et~al.}()\citenamefont {Yu},
  \citenamefont {Novikov}, \citenamefont {Ting}, \citenamefont {Sarney},
  \citenamefont {Svensson}, \citenamefont {Shaw}, \citenamefont {Martin},
  \citenamefont {Walukiewicz},\ and\ \citenamefont {Foxon}}]{YuNo2014}%
  \BibitemOpen
  \bibfield  {author} {\bibinfo {author} {\bibfnamefont {K.~M.}\ \bibnamefont
  {Yu}}, \bibinfo {author} {\bibfnamefont {S.~V.}\ \bibnamefont {Novikov}},
  \bibinfo {author} {\bibfnamefont {M.}~\bibnamefont {Ting}}, \bibinfo {author}
  {\bibfnamefont {W.~L.}\ \bibnamefont {Sarney}}, \bibinfo {author}
  {\bibfnamefont {S.~P.}\ \bibnamefont {Svensson}}, \bibinfo {author}
  {\bibfnamefont {M.}~\bibnamefont {Shaw}}, \bibinfo {author} {\bibfnamefont
  {R.~W.}\ \bibnamefont {Martin}}, \bibinfo {author} {\bibfnamefont
  {W.}~\bibnamefont {Walukiewicz}},\ and\ \bibinfo {author} {\bibfnamefont
  {C.~T.}\ \bibnamefont {Foxon}},\ }\bibfield  {title} {\bibinfo {title}
  {Growth and characterization of highly mismatched {GaN}$_{1-x}${Sb}$_{x}$
  alloys},\ }\href {https://doi.org/10.1063/1.4896364} {\bibfield  {journal}
  {\bibinfo  {journal} {Journal of Applied Physics}\ }\textbf {\bibinfo
  {volume} {116}},\ \bibinfo {pages} {123704}}\BibitemShut {NoStop}%
\bibitem [{\citenamefont {Chowdhury}\ and\ \citenamefont {Mi}()}]{ChMi2019}%
  \BibitemOpen
  \bibfield  {author} {\bibinfo {author} {\bibfnamefont {F.~A.}\ \bibnamefont
  {Chowdhury}}\ and\ \bibinfo {author} {\bibfnamefont {Z.}~\bibnamefont {Mi}},\
  }\bibfield  {title} {\bibinfo {title} {Probing the large bandgap-bowing and
  signature of antimony (sb) in dilute-antimonide {III}-nitride using
  micro-raman scattering},\ }\href {https://doi.org/10.1063/1.5109735}
  {\bibfield  {journal} {\bibinfo  {journal} {Journal of Applied Physics}\
  }\textbf {\bibinfo {volume} {126}},\ \bibinfo {pages} {085704}}\BibitemShut
  {NoStop}%
\bibitem [{\citenamefont {Nye}(1985)}]{Ny1985}%
  \BibitemOpen
  \bibfield  {author} {\bibinfo {author} {\bibfnamefont {J.~F.}\ \bibnamefont
  {Nye}},\ }\href@noop {} {\emph {\bibinfo {title} {Physicsal Properties of
  Crystals: Their Representation by Tensors and Matrices.}}}\ (\bibinfo
  {publisher} {Oxford University Press},\ \bibinfo {year} {1985})\BibitemShut
  {NoStop}%
\bibitem [{\citenamefont {Caro}\ \emph
  {et~al.}(2012{\natexlab{a}})\citenamefont {Caro}, \citenamefont {Schulz},\
  and\ \citenamefont {O'Reilly}}]{Caro2012}%
  \BibitemOpen
  \bibfield  {author} {\bibinfo {author} {\bibfnamefont {M.~A.}\ \bibnamefont
  {Caro}}, \bibinfo {author} {\bibfnamefont {S.}~\bibnamefont {Schulz}},\ and\
  \bibinfo {author} {\bibfnamefont {E.~P.}\ \bibnamefont {O'Reilly}},\
  }\bibfield  {title} {\bibinfo {title} {Hybrid functional study of the elastic
  and structural properties of wurtzite and zinc-blende group-iii nitrides},\
  }\href {https://doi.org/10.1103/PhysRevB.86.014117} {\bibfield  {journal}
  {\bibinfo  {journal} {Phys. Rev. B}\ }\textbf {\bibinfo {volume} {86}},\
  \bibinfo {pages} {014117} (\bibinfo {year} {2012}{\natexlab{a}})}\BibitemShut
  {NoStop}%
\bibitem [{\citenamefont {Binks}\ \emph {et~al.}()\citenamefont {Binks},
  \citenamefont {Dawson}, \citenamefont {Oliver},\ and\ \citenamefont
  {Wallis}}]{BiDa2022}%
  \BibitemOpen
  \bibfield  {author} {\bibinfo {author} {\bibfnamefont {D.~J.}\ \bibnamefont
  {Binks}}, \bibinfo {author} {\bibfnamefont {P.}~\bibnamefont {Dawson}},
  \bibinfo {author} {\bibfnamefont {R.~A.}\ \bibnamefont {Oliver}},\ and\
  \bibinfo {author} {\bibfnamefont {D.~J.}\ \bibnamefont {Wallis}},\ }\bibfield
   {title} {\bibinfo {title} {Cubic {GaN} and {InGaN}/{GaN} quantum wells},\
  }\href {https://doi.org/10.1063/5.0097558} {\bibfield  {journal} {\bibinfo
  {journal} {Applied Physics Reviews}\ }\textbf {\bibinfo {volume} {9}},\
  \bibinfo {pages} {041309}}\BibitemShut {NoStop}%
\bibitem [{\citenamefont {Grosse}\ and\ \citenamefont
  {Neugebauer}(2001)}]{GrNe2001}%
  \BibitemOpen
  \bibfield  {author} {\bibinfo {author} {\bibfnamefont {F.}~\bibnamefont
  {Grosse}}\ and\ \bibinfo {author} {\bibfnamefont {J.}~\bibnamefont
  {Neugebauer}},\ }\bibfield  {title} {\bibinfo {title} {Limits and accuracy of
  valence force field models for {In$_x$Ga$_1-x$N} alloys},\ }\href
  {https://doi.org/10.1103/PhysRevB.63.085207} {\bibfield  {journal} {\bibinfo
  {journal} {Phys. Rev. B}\ }\textbf {\bibinfo {volume} {63}},\ \bibinfo
  {pages} {085207} (\bibinfo {year} {2001})}\BibitemShut {NoStop}%
\bibitem [{\citenamefont {Camacho}\ and\ \citenamefont
  {Niquet}(2010)}]{CaNi2010}%
  \BibitemOpen
  \bibfield  {author} {\bibinfo {author} {\bibfnamefont {D.}~\bibnamefont
  {Camacho}}\ and\ \bibinfo {author} {\bibfnamefont {Y.}~\bibnamefont
  {Niquet}},\ }\bibfield  {title} {\bibinfo {title} {Application of keating's
  valence force field model to non-ideal wurtzite materials},\ }\href
  {https://doi.org/https://doi.org/10.1016/j.physe.2009.11.035} {\bibfield
  {journal} {\bibinfo  {journal} {Physica E: Low-dimensional Systems and
  Nanostructures}\ }\textbf {\bibinfo {volume} {42}},\ \bibinfo {pages} {1361}
  (\bibinfo {year} {2010})}\BibitemShut {NoStop}%
\bibitem [{\citenamefont {Musgrave}\ and\ \citenamefont
  {Pople}(1962)}]{MuPo1962}%
  \BibitemOpen
  \bibfield  {author} {\bibinfo {author} {\bibfnamefont {M.~J.~P.}\
  \bibnamefont {Musgrave}}\ and\ \bibinfo {author} {\bibfnamefont {J.~A.}\
  \bibnamefont {Pople}},\ }\bibfield  {title} {\bibinfo {title} {A general
  valence force field for diamond},\ }\href
  {https://doi.org/10.1098/rspa.1962.0153} {\bibfield  {journal} {\bibinfo
  {journal} {Proc. R. Soc. London A}\ }\textbf {\bibinfo {volume} {268}},\
  \bibinfo {pages} {474–484} (\bibinfo {year} {1962})}\BibitemShut {NoStop}%
\bibitem [{\citenamefont {Martin}(1970)}]{Mart1970}%
  \BibitemOpen
  \bibfield  {author} {\bibinfo {author} {\bibfnamefont {R.~M.}\ \bibnamefont
  {Martin}},\ }\bibfield  {title} {\bibinfo {title} {Elastic properties of zns
  structure semiconductors},\ }\href {https://doi.org/10.1103/PhysRevB.1.4005}
  {\bibfield  {journal} {\bibinfo  {journal} {Phys. Rev. B}\ }\textbf {\bibinfo
  {volume} {1}},\ \bibinfo {pages} {4005} (\bibinfo {year} {1970})}\BibitemShut
  {NoStop}%
\bibitem [{\citenamefont {Blackman}(1958)}]{Blac1958}%
  \BibitemOpen
  \bibfield  {author} {\bibinfo {author} {\bibfnamefont {M.}~\bibnamefont
  {Blackman}},\ }\bibfield  {title} {\bibinfo {title} {On negative volume
  expansion coefficients},\ }\href {https://doi.org/10.1080/14786435808237021}
  {\bibfield  {journal} {\bibinfo  {journal} {The Philosophical Magazine: A
  Journal of Theoretical Experimental and Applied Physics}\ }\textbf {\bibinfo
  {volume} {3}},\ \bibinfo {pages} {831} (\bibinfo {year} {1958})}\BibitemShut
  {NoStop}%
\bibitem [{\citenamefont {Schulz}\ \emph {et~al.}(2011)\citenamefont {Schulz},
  \citenamefont {Caro}, \citenamefont {O'Reilly},\ and\ \citenamefont
  {Marquardt}}]{ScCa2011}%
  \BibitemOpen
  \bibfield  {author} {\bibinfo {author} {\bibfnamefont {S.}~\bibnamefont
  {Schulz}}, \bibinfo {author} {\bibfnamefont {M.~A.}\ \bibnamefont {Caro}},
  \bibinfo {author} {\bibfnamefont {E.~P.}\ \bibnamefont {O'Reilly}},\ and\
  \bibinfo {author} {\bibfnamefont {O.}~\bibnamefont {Marquardt}},\ }\bibfield
  {title} {\bibinfo {title} {Symmetry-adapted calculations of strain and
  polarization fields in (111)-oriented zinc-blende quantum dots},\ }\href
  {https://doi.org/10.1103/PhysRevB.84.125312} {\bibfield  {journal} {\bibinfo
  {journal} {Phys. Rev. B}\ }\textbf {\bibinfo {volume} {84}},\ \bibinfo
  {pages} {125312} (\bibinfo {year} {2011})}\BibitemShut {NoStop}%
\bibitem [{\citenamefont {Humphreys}(2008)}]{HumphreysSolidStateLighting}%
  \BibitemOpen
  \bibfield  {author} {\bibinfo {author} {\bibfnamefont {C.~J.}\ \bibnamefont
  {Humphreys}},\ }\bibfield  {title} {\bibinfo {title} {Solid-state lighting},\
  }\href {https://doi.org/10.1557/mrs2008.91} {\bibfield  {journal} {\bibinfo
  {journal} {MRS Bulletin}\ }\textbf {\bibinfo {volume} {33}},\ \bibinfo
  {pages} {459–470} (\bibinfo {year} {2008})}\BibitemShut {NoStop}%
\bibitem [{\citenamefont {Nelder}\ and\ \citenamefont {Mead}(1965)}]{NeMe1965}%
  \BibitemOpen
  \bibfield  {author} {\bibinfo {author} {\bibfnamefont {J.~A.}\ \bibnamefont
  {Nelder}}\ and\ \bibinfo {author} {\bibfnamefont {R.}~\bibnamefont {Mead}},\
  }\bibfield  {title} {\bibinfo {title} {{A Simplex Method for Function
  Minimization}},\ }\href {https://doi.org/10.1093/comjnl/7.4.308} {\bibfield
  {journal} {\bibinfo  {journal} {The Computer Journal}\ }\textbf {\bibinfo
  {volume} {7}},\ \bibinfo {pages} {308} (\bibinfo {year} {1965})},\ \Eprint
  {https://arxiv.org/abs/https://academic.oup.com/comjnl/article-pdf/7/4/308/1013182/7-4-308.pdf}
  {https://academic.oup.com/comjnl/article-pdf/7/4/308/1013182/7-4-308.pdf}
  \BibitemShut {NoStop}%
\bibitem [{\citenamefont {Bezanson}\ \emph {et~al.}(2017)\citenamefont
  {Bezanson}, \citenamefont {Edelman}, \citenamefont {Karpinski},\ and\
  \citenamefont {Shah}}]{Julia2017}%
  \BibitemOpen
  \bibfield  {author} {\bibinfo {author} {\bibfnamefont {J.}~\bibnamefont
  {Bezanson}}, \bibinfo {author} {\bibfnamefont {A.}~\bibnamefont {Edelman}},
  \bibinfo {author} {\bibfnamefont {S.}~\bibnamefont {Karpinski}},\ and\
  \bibinfo {author} {\bibfnamefont {V.~B.}\ \bibnamefont {Shah}},\ }\bibfield
  {title} {\bibinfo {title} {Julia: A fresh approach to numerical computing},\
  }\href {https://doi.org/10.1137/141000671} {\bibfield  {journal} {\bibinfo
  {journal} {SIAM Review}\ }\textbf {\bibinfo {volume} {59}},\ \bibinfo {pages}
  {65} (\bibinfo {year} {2017})}\BibitemShut {NoStop}%
\bibitem [{\citenamefont {Mogensen}\ and\ \citenamefont
  {Riseth}(2018)}]{OptimJL2018}%
  \BibitemOpen
  \bibfield  {author} {\bibinfo {author} {\bibfnamefont {P.~K.}\ \bibnamefont
  {Mogensen}}\ and\ \bibinfo {author} {\bibfnamefont {A.~N.}\ \bibnamefont
  {Riseth}},\ }\bibfield  {title} {\bibinfo {title} {Optim: A mathematical
  optimization package for julia},\ }\href
  {https://doi.org/10.21105/JOSS.00615} {\bibfield  {journal} {\bibinfo
  {journal} {Journal of Open Source Software}\ }\textbf {\bibinfo {volume}
  {3}},\ \bibinfo {pages} {615} (\bibinfo {year} {2018})}\BibitemShut {NoStop}%
\bibitem [{\citenamefont {Caro}\ \emph {et~al.}(2013)\citenamefont {Caro},
  \citenamefont {Schulz},\ and\ \citenamefont {O'Reilly}}]{CaSc2013}%
  \BibitemOpen
  \bibfield  {author} {\bibinfo {author} {\bibfnamefont {M.~A.}\ \bibnamefont
  {Caro}}, \bibinfo {author} {\bibfnamefont {S.}~\bibnamefont {Schulz}},\ and\
  \bibinfo {author} {\bibfnamefont {E.~P.}\ \bibnamefont {O'Reilly}},\
  }\bibfield  {title} {\bibinfo {title} {Theory of local electric polarization
  and its relation to internal strain: Impact on polarization potential and
  electronic properties of group-iii nitrides},\ }\href
  {https://doi.org/10.1103/PhysRevB.88.214103} {\bibfield  {journal} {\bibinfo
  {journal} {Phys. Rev. B}\ }\textbf {\bibinfo {volume} {88}},\ \bibinfo
  {pages} {214103} (\bibinfo {year} {2013})}\BibitemShut {NoStop}%
\bibitem [{\citenamefont {Mattila}\ and\ \citenamefont
  {Zunger}(1998)}]{MaZu1998}%
  \BibitemOpen
  \bibfield  {author} {\bibinfo {author} {\bibfnamefont {T.}~\bibnamefont
  {Mattila}}\ and\ \bibinfo {author} {\bibfnamefont {A.}~\bibnamefont
  {Zunger}},\ }\bibfield  {title} {\bibinfo {title} {Predicted bond length
  variation in wurtzite and zinc-blende ingan and algan alloys},\ }\href
  {https://doi.org/10.1063/1.369463} {\bibfield  {journal} {\bibinfo  {journal}
  {Appl. Phys. Lett.}\ }\textbf {\bibinfo {volume} {85}},\ \bibinfo {pages}
  {160} (\bibinfo {year} {1998})}\BibitemShut {NoStop}%
\bibitem [{\citenamefont {Gale}(1997)}]{GULP1997}%
  \BibitemOpen
  \bibfield  {author} {\bibinfo {author} {\bibfnamefont {J.~D.}\ \bibnamefont
  {Gale}},\ }\bibfield  {title} {\bibinfo {title} {Gulp: A computer program for
  the symmetry-adapted simulation of solids},\ }\href
  {https://doi.org/10.1039/A606455H} {\bibfield  {journal} {\bibinfo  {journal}
  {J. Chem. Soc.{,} Faraday Trans.}\ }\textbf {\bibinfo {volume} {93}},\
  \bibinfo {pages} {629} (\bibinfo {year} {1997})}\BibitemShut {NoStop}%
\bibitem [{\citenamefont {Gale}\ and\ \citenamefont {Rohl}(2003)}]{GULP2003}%
  \BibitemOpen
  \bibfield  {author} {\bibinfo {author} {\bibfnamefont {J.~D.}\ \bibnamefont
  {Gale}}\ and\ \bibinfo {author} {\bibfnamefont {A.~L.}\ \bibnamefont
  {Rohl}},\ }\bibfield  {title} {\bibinfo {title} {The general utility lattice
  program (gulp)},\ }\href {https://doi.org/10.1080/0892702031000104887}
  {\bibfield  {journal} {\bibinfo  {journal} {Molecular Simulation}\ }\textbf
  {\bibinfo {volume} {29}},\ \bibinfo {pages} {291} (\bibinfo {year}
  {2003})}\BibitemShut {NoStop}%
\bibitem [{\citenamefont {Sheerin}\ \emph {et~al.}()\citenamefont {Sheerin},
  \citenamefont {Tanner},\ and\ \citenamefont {Schulz}}]{ShTa2021}%
  \BibitemOpen
  \bibfield  {author} {\bibinfo {author} {\bibfnamefont {T.~P.}\ \bibnamefont
  {Sheerin}}, \bibinfo {author} {\bibfnamefont {D.~S.~P.}\ \bibnamefont
  {Tanner}},\ and\ \bibinfo {author} {\bibfnamefont {S.}~\bibnamefont
  {Schulz}},\ }\bibfield  {title} {\bibinfo {title} {Atomistic analysis of
  piezoelectric potential fluctuations in zinc-blende {InGaN}/{GaN} quantum
  wells: A stillinger-weber potential based analysis},\ }\href
  {https://link.aps.org/doi/10.1103/PhysRevB.103.165201} {\bibfield  {journal}
  {\bibinfo  {journal} {Physical Review B}\ }\textbf {\bibinfo {volume}
  {103}},\ \bibinfo {pages} {165201}}\BibitemShut {NoStop}%
\bibitem [{\citenamefont {Klimeck}\ \emph {et~al.}()\citenamefont {Klimeck},
  \citenamefont {Ahmed}, \citenamefont {Bae}, \citenamefont {Kharche},
  \citenamefont {Clark}, \citenamefont {Haley}, \citenamefont {Lee},
  \citenamefont {Naumov}, \citenamefont {Ryu}, \citenamefont {Saied},
  \citenamefont {Prada}, \citenamefont {Korkusinski}, \citenamefont {Boykin},\
  and\ \citenamefont {Rahman}}]{NEMO2007}%
  \BibitemOpen
  \bibfield  {author} {\bibinfo {author} {\bibfnamefont {G.}~\bibnamefont
  {Klimeck}}, \bibinfo {author} {\bibfnamefont {S.}~\bibnamefont {Ahmed}},
  \bibinfo {author} {\bibfnamefont {H.}~\bibnamefont {Bae}}, \bibinfo {author}
  {\bibfnamefont {N.}~\bibnamefont {Kharche}}, \bibinfo {author} {\bibfnamefont
  {S.}~\bibnamefont {Clark}}, \bibinfo {author} {\bibfnamefont
  {B.}~\bibnamefont {Haley}}, \bibinfo {author} {\bibfnamefont
  {S.}~\bibnamefont {Lee}}, \bibinfo {author} {\bibfnamefont {M.}~\bibnamefont
  {Naumov}}, \bibinfo {author} {\bibfnamefont {H.}~\bibnamefont {Ryu}},
  \bibinfo {author} {\bibfnamefont {F.}~\bibnamefont {Saied}}, \bibinfo
  {author} {\bibfnamefont {M.}~\bibnamefont {Prada}}, \bibinfo {author}
  {\bibfnamefont {M.}~\bibnamefont {Korkusinski}}, \bibinfo {author}
  {\bibfnamefont {T.}~\bibnamefont {Boykin}},\ and\ \bibinfo {author}
  {\bibfnamefont {R.}~\bibnamefont {Rahman}},\ }\bibfield  {title} {\bibinfo
  {title} {Atomistic simulation of realistically sized nanodevices using {NEMO}
  3-d—part i: Models and benchmarks},\ }\href
  {https://doi.org/10.1109/TED.2007.902879} {\bibfield  {journal} {\bibinfo
  {journal} {{IEEE} Transactions on Electron Devices}\ }\textbf {\bibinfo
  {volume} {54}},\ \bibinfo {pages} {2079}}\BibitemShut {NoStop}%
\bibitem [{\citenamefont {Sheerin}\ and\ \citenamefont
  {Schulz}(2022)}]{Sheerin2022}%
  \BibitemOpen
  \bibfield  {author} {\bibinfo {author} {\bibfnamefont {T.~P.}\ \bibnamefont
  {Sheerin}}\ and\ \bibinfo {author} {\bibfnamefont {S.}~\bibnamefont
  {Schulz}},\ }\bibfield  {title} {\bibinfo {title} {Strain effects in wurtzite
  boron nitride: Elastic constants, internal strain, and deformation potentials
  from hybrid functional density functional theory},\ }\href
  {https://doi.org/https://doi.org/10.1002/pssr.202200021} {\bibfield
  {journal} {\bibinfo  {journal} {physica status solidi (RRL) – Rapid
  Research Letters}\ }\textbf {\bibinfo {volume} {16}},\ \bibinfo {pages}
  {2200021} (\bibinfo {year} {2022})}\BibitemShut {NoStop}%
\bibitem [{\citenamefont {Yan}\ \emph {et~al.}(2011)\citenamefont {Yan},
  \citenamefont {Rinke}, \citenamefont {Winkelnkemper}, \citenamefont {Qteish},
  \citenamefont {Bimberg}, \citenamefont {Scheffler},\ and\ \citenamefont {{C.
  G. Van de Walle}}}]{YaRi2011}%
  \BibitemOpen
  \bibfield  {author} {\bibinfo {author} {\bibfnamefont {Q.}~\bibnamefont
  {Yan}}, \bibinfo {author} {\bibfnamefont {P.}~\bibnamefont {Rinke}}, \bibinfo
  {author} {\bibfnamefont {M.}~\bibnamefont {Winkelnkemper}}, \bibinfo {author}
  {\bibfnamefont {A.}~\bibnamefont {Qteish}}, \bibinfo {author} {\bibfnamefont
  {D.}~\bibnamefont {Bimberg}}, \bibinfo {author} {\bibfnamefont
  {M.}~\bibnamefont {Scheffler}},\ and\ \bibinfo {author} {\bibnamefont {{C. G.
  Van de Walle}}},\ }\bibfield  {title} {\bibinfo {title} {Band parameters and
  strain effects in {ZnO} and {group-III} nitrides},\ }\href
  {https://doi.org/10.1088/0268-1242/26/1/014037} {\bibfield  {journal}
  {\bibinfo  {journal} {Semicond. Sci. Technol.}\ }\textbf {\bibinfo {volume}
  {26}},\ \bibinfo {pages} {014037} (\bibinfo {year} {2011})}\BibitemShut
  {NoStop}%
\bibitem [{\citenamefont {Kresse}\ and\ \citenamefont
  {Furthm\"uller}(1996)}]{VASP}%
  \BibitemOpen
  \bibfield  {author} {\bibinfo {author} {\bibfnamefont {G.}~\bibnamefont
  {Kresse}}\ and\ \bibinfo {author} {\bibfnamefont {J.}~\bibnamefont
  {Furthm\"uller}},\ }\bibfield  {title} {\bibinfo {title} {Efficient iterative
  schemes for ab initio total-energy calculations using a plane-wave basis
  set},\ }\href {https://doi.org/10.1103/PhysRevB.54.11169} {\bibfield
  {journal} {\bibinfo  {journal} {Phys. Rev. B}\ }\textbf {\bibinfo {volume}
  {54}},\ \bibinfo {pages} {11169} (\bibinfo {year} {1996})}\BibitemShut
  {NoStop}%
\bibitem [{\citenamefont {Xiao}\ \emph {et~al.}(2011)\citenamefont {Xiao},
  \citenamefont {Tahir-Kheli},\ and\ \citenamefont {Goddard~III}}]{xiao11}%
  \BibitemOpen
  \bibfield  {author} {\bibinfo {author} {\bibfnamefont {H.}~\bibnamefont
  {Xiao}}, \bibinfo {author} {\bibfnamefont {J.}~\bibnamefont {Tahir-Kheli}},\
  and\ \bibinfo {author} {\bibfnamefont {W.~A.}\ \bibnamefont {Goddard~III}},\
  }\bibfield  {title} {\bibinfo {title} {Accurate band gaps for semiconductors
  from density functional theory},\ }\href@noop {} {\bibfield  {journal}
  {\bibinfo  {journal} {The Journal of Physical Chemistry Letters}\ }\textbf
  {\bibinfo {volume} {2}},\ \bibinfo {pages} {212} (\bibinfo {year}
  {2011})}\BibitemShut {NoStop}%
\bibitem [{\citenamefont {Perdew}\ \emph {et~al.}(2017)\citenamefont {Perdew},
  \citenamefont {Yang}, \citenamefont {Burke}, \citenamefont {Yang},
  \citenamefont {Gross}, \citenamefont {Scheffler}, \citenamefont {Scuseria},
  \citenamefont {Henderson}, \citenamefont {Zhang}, \citenamefont {Ruzsinszky}
  \emph {et~al.}}]{perdew17}%
  \BibitemOpen
  \bibfield  {author} {\bibinfo {author} {\bibfnamefont {J.~P.}\ \bibnamefont
  {Perdew}}, \bibinfo {author} {\bibfnamefont {W.}~\bibnamefont {Yang}},
  \bibinfo {author} {\bibfnamefont {K.}~\bibnamefont {Burke}}, \bibinfo
  {author} {\bibfnamefont {Z.}~\bibnamefont {Yang}}, \bibinfo {author}
  {\bibfnamefont {E.~K.}\ \bibnamefont {Gross}}, \bibinfo {author}
  {\bibfnamefont {M.}~\bibnamefont {Scheffler}}, \bibinfo {author}
  {\bibfnamefont {G.~E.}\ \bibnamefont {Scuseria}}, \bibinfo {author}
  {\bibfnamefont {T.~M.}\ \bibnamefont {Henderson}}, \bibinfo {author}
  {\bibfnamefont {I.~Y.}\ \bibnamefont {Zhang}}, \bibinfo {author}
  {\bibfnamefont {A.}~\bibnamefont {Ruzsinszky}}, \emph {et~al.},\ }\bibfield
  {title} {\bibinfo {title} {Understanding band gaps of solids in generalized
  kohn--sham theory},\ }\href@noop {} {\bibfield  {journal} {\bibinfo
  {journal} {Proceedings of the national academy of sciences}\ }\textbf
  {\bibinfo {volume} {114}},\ \bibinfo {pages} {2801} (\bibinfo {year}
  {2017})}\BibitemShut {NoStop}%
\bibitem [{\citenamefont {Lany}\ and\ \citenamefont {Zunger}(2008)}]{lany08}%
  \BibitemOpen
  \bibfield  {author} {\bibinfo {author} {\bibfnamefont {S.}~\bibnamefont
  {Lany}}\ and\ \bibinfo {author} {\bibfnamefont {A.}~\bibnamefont {Zunger}},\
  }\bibfield  {title} {\bibinfo {title} {Assessment of correction methods for
  the band-gap problem and for finite-size effects in supercell defect
  calculations: Case studies for zno and gaas},\ }\href@noop {} {\bibfield
  {journal} {\bibinfo  {journal} {Physical Review B}\ }\textbf {\bibinfo
  {volume} {78}},\ \bibinfo {pages} {235104} (\bibinfo {year}
  {2008})}\BibitemShut {NoStop}%
\bibitem [{\citenamefont {Heyd}\ \emph {et~al.}(2003)\citenamefont {Heyd},
  \citenamefont {Scuseria},\ and\ \citenamefont {Ernzerhof}}]{HSEDFT}%
  \BibitemOpen
  \bibfield  {author} {\bibinfo {author} {\bibfnamefont {J.}~\bibnamefont
  {Heyd}}, \bibinfo {author} {\bibfnamefont {G.~E.}\ \bibnamefont {Scuseria}},\
  and\ \bibinfo {author} {\bibfnamefont {M.}~\bibnamefont {Ernzerhof}},\
  }\bibfield  {title} {\bibinfo {title} {{Hybrid functionals based on a
  screened Coulomb potential}},\ }\href {https://doi.org/10.1063/1.1564060}
  {\bibfield  {journal} {\bibinfo  {journal} {The Journal of Chemical Physics}\
  }\textbf {\bibinfo {volume} {118}},\ \bibinfo {pages} {8207} (\bibinfo {year}
  {2003})}\BibitemShut {NoStop}%
\bibitem [{\citenamefont {Krukau}\ \emph {et~al.}(2006)\citenamefont {Krukau},
  \citenamefont {Vydrov}, \citenamefont {Izmaylov},\ and\ \citenamefont
  {Scuseria}}]{Krukau2006_HSE}%
  \BibitemOpen
  \bibfield  {author} {\bibinfo {author} {\bibfnamefont {A.~V.}\ \bibnamefont
  {Krukau}}, \bibinfo {author} {\bibfnamefont {O.~A.}\ \bibnamefont {Vydrov}},
  \bibinfo {author} {\bibfnamefont {A.~F.}\ \bibnamefont {Izmaylov}},\ and\
  \bibinfo {author} {\bibfnamefont {G.~E.}\ \bibnamefont {Scuseria}},\
  }\bibfield  {title} {\bibinfo {title} {{Influence of the exchange screening
  parameter on the performance of screened hybrid functionals}},\ }\href
  {https://doi.org/10.1063/1.2404663} {\bibfield  {journal} {\bibinfo
  {journal} {The Journal of Chemical Physics}\ }\textbf {\bibinfo {volume}
  {125}},\ \bibinfo {pages} {224106} (\bibinfo {year} {2006})},\ \Eprint
  {https://arxiv.org/abs/https://pubs.aip.org/aip/jcp/article-pdf/doi/10.1063/1.2404663/13263224/224106\_1\_online.pdf}
  {https://pubs.aip.org/aip/jcp/article-pdf/doi/10.1063/1.2404663/13263224/224106\_1\_online.pdf}
  \BibitemShut {NoStop}%
\bibitem [{\citenamefont {Borlido}\ \emph {et~al.}(2020)\citenamefont
  {Borlido}, \citenamefont {Schmidt}, \citenamefont {Huran}, \citenamefont
  {Tran}, \citenamefont {Marques},\ and\ \citenamefont {Botti}}]{Borlido2020}%
  \BibitemOpen
  \bibfield  {author} {\bibinfo {author} {\bibfnamefont {P.}~\bibnamefont
  {Borlido}}, \bibinfo {author} {\bibfnamefont {J.}~\bibnamefont {Schmidt}},
  \bibinfo {author} {\bibfnamefont {A.~W.}\ \bibnamefont {Huran}}, \bibinfo
  {author} {\bibfnamefont {F.}~\bibnamefont {Tran}}, \bibinfo {author}
  {\bibfnamefont {M.~A.~L.}\ \bibnamefont {Marques}},\ and\ \bibinfo {author}
  {\bibfnamefont {S.}~\bibnamefont {Botti}},\ }\bibfield  {title} {\bibinfo
  {title} {Exchange-correlation functionals for band gaps of solids: benchmark,
  reparametrization and machine learning},\ }\href
  {https://doi.org/10.1038/s41524-020-00360-0} {\bibfield  {journal} {\bibinfo
  {journal} {npj Computational Materials}\ }\textbf {\bibinfo {volume} {6}},\
  \bibinfo {pages} {96} (\bibinfo {year} {2020})}\BibitemShut {NoStop}%
\bibitem [{\citenamefont {Becke}\ and\ \citenamefont {Johnson}(2006)}]{mBJ}%
  \BibitemOpen
  \bibfield  {author} {\bibinfo {author} {\bibfnamefont {A.~D.}\ \bibnamefont
  {Becke}}\ and\ \bibinfo {author} {\bibfnamefont {E.~R.}\ \bibnamefont
  {Johnson}},\ }\bibfield  {title} {\bibinfo {title} {{A simple effective
  potential for exchange}},\ }\href {https://doi.org/10.1063/1.2213970}
  {\bibfield  {journal} {\bibinfo  {journal} {The Journal of Chemical Physics}\
  }\textbf {\bibinfo {volume} {124}},\ \bibinfo {pages} {221101} (\bibinfo
  {year} {2006})},\ \Eprint
  {https://arxiv.org/abs/https://pubs.aip.org/aip/jcp/article-pdf/doi/10.1063/1.2213970/15385734/221101\_1\_online.pdf}
  {https://pubs.aip.org/aip/jcp/article-pdf/doi/10.1063/1.2213970/15385734/221101\_1\_online.pdf}
  \BibitemShut {NoStop}%
\bibitem [{\citenamefont {Tran}\ and\ \citenamefont {Blaha}(2009)}]{Tran2009}%
  \BibitemOpen
  \bibfield  {author} {\bibinfo {author} {\bibfnamefont {F.}~\bibnamefont
  {Tran}}\ and\ \bibinfo {author} {\bibfnamefont {P.}~\bibnamefont {Blaha}},\
  }\bibfield  {title} {\bibinfo {title} {Accurate band gaps of semiconductors
  and insulators with a semilocal exchange-correlation potential},\ }\href
  {https://doi.org/10.1103/PhysRevLett.102.226401} {\bibfield  {journal}
  {\bibinfo  {journal} {Phys. Rev. Lett.}\ }\textbf {\bibinfo {volume} {102}},\
  \bibinfo {pages} {226401} (\bibinfo {year} {2009})}\BibitemShut {NoStop}%
\bibitem [{\citenamefont {Perdew}\ \emph {et~al.}(1996)\citenamefont {Perdew},
  \citenamefont {Burke},\ and\ \citenamefont {Ernzerhof}}]{PBE}%
  \BibitemOpen
  \bibfield  {author} {\bibinfo {author} {\bibfnamefont {J.~P.}\ \bibnamefont
  {Perdew}}, \bibinfo {author} {\bibfnamefont {K.}~\bibnamefont {Burke}},\ and\
  \bibinfo {author} {\bibfnamefont {M.}~\bibnamefont {Ernzerhof}},\ }\bibfield
  {title} {\bibinfo {title} {Generalized gradient approximation made simple},\
  }\href {https://doi.org/10.1103/PhysRevLett.77.3865} {\bibfield  {journal}
  {\bibinfo  {journal} {Phys. Rev. Lett.}\ }\textbf {\bibinfo {volume} {77}},\
  \bibinfo {pages} {3865} (\bibinfo {year} {1996})}\BibitemShut {NoStop}%
\bibitem [{\citenamefont {Koch}\ and\ \citenamefont
  {Holthausen}(2001)}]{koch01-4}%
  \BibitemOpen
  \bibfield  {author} {\bibinfo {author} {\bibfnamefont {W.}~\bibnamefont
  {Koch}}\ and\ \bibinfo {author} {\bibfnamefont {M.}~\bibnamefont
  {Holthausen}},\ }\href@noop {} {\emph {\bibinfo {title} {A Chemist’s Guide
  to Density Functional Theory, Wiley}}}\ (\bibinfo  {publisher} {VCH,
  Weinheim, Germany},\ \bibinfo {year} {2001})\ pp.\ \bibinfo {pages}
  {65--91}\BibitemShut {NoStop}%
\bibitem [{\citenamefont {Caro}\ \emph
  {et~al.}(2012{\natexlab{b}})\citenamefont {Caro}, \citenamefont {Schulz},\
  and\ \citenamefont {O’Reilly}}]{CaSc2013_2}%
  \BibitemOpen
  \bibfield  {author} {\bibinfo {author} {\bibfnamefont {M.~A.}\ \bibnamefont
  {Caro}}, \bibinfo {author} {\bibfnamefont {S.}~\bibnamefont {Schulz}},\ and\
  \bibinfo {author} {\bibfnamefont {E.~P.}\ \bibnamefont {O’Reilly}},\
  }\bibfield  {title} {\bibinfo {title} {Comparison of stress and total energy
  methods for calculation of elastic properties of semiconductors},\ }\href
  {https://doi.org/10.1088/0953-8984/25/2/025803} {\bibfield  {journal}
  {\bibinfo  {journal} {Journal of Physics: Condensed Matter}\ }\textbf
  {\bibinfo {volume} {25}},\ \bibinfo {pages} {025803} (\bibinfo {year}
  {2012}{\natexlab{b}})}\BibitemShut {NoStop}%
\bibitem [{\citenamefont {Nies}\ \emph {et~al.}(2023)\citenamefont {Nies},
  \citenamefont {Sheerin},\ and\ \citenamefont {Schulz}}]{NiSh2023}%
  \BibitemOpen
  \bibfield  {author} {\bibinfo {author} {\bibfnamefont {C.~L.}\ \bibnamefont
  {Nies}}, \bibinfo {author} {\bibfnamefont {T.~P.}\ \bibnamefont {Sheerin}},\
  and\ \bibinfo {author} {\bibfnamefont {S.}~\bibnamefont {Schulz}},\
  }\bibfield  {title} {\bibinfo {title} {Electronic and optical properties of
  boron-containing gan alloys: The role of boron atom clustering},\ }\href
  {https://doi.org/10.1063/5.0171932/2912943} {\bibfield  {journal} {\bibinfo
  {journal} {APL Materials}\ }\textbf {\bibinfo {volume} {11}},\ \bibinfo
  {pages} {91119} (\bibinfo {year} {2023})}\BibitemShut {NoStop}%
\bibitem [{\citenamefont {Williams}\ and\ \citenamefont
  {Kioupakis}(2017)}]{LoKi2017}%
  \BibitemOpen
  \bibfield  {author} {\bibinfo {author} {\bibfnamefont {L.}~\bibnamefont
  {Williams}}\ and\ \bibinfo {author} {\bibfnamefont {E.}~\bibnamefont
  {Kioupakis}},\ }\bibfield  {title} {\bibinfo {title} {{BInGaN alloys nearly
  lattice-matched to GaN for high-power high-efficiency visible LEDs}},\ }\href
  {https://doi.org/10.1063/1.4997601} {\bibfield  {journal} {\bibinfo
  {journal} {Applied Physics Letters}\ }\textbf {\bibinfo {volume} {111}},\
  \bibinfo {pages} {211107} (\bibinfo {year} {2017})},\ \Eprint
  {https://arxiv.org/abs/https://pubs.aip.org/aip/apl/article-pdf/doi/10.1063/1.4997601/13316240/211107\_1\_online.pdf}
  {https://pubs.aip.org/aip/apl/article-pdf/doi/10.1063/1.4997601/13316240/211107\_1\_online.pdf}
  \BibitemShut {NoStop}%
\bibitem [{\citenamefont {Rasander}\ and\ \citenamefont
  {Moram}(2015)}]{RaMo2015}%
  \BibitemOpen
  \bibfield  {author} {\bibinfo {author} {\bibfnamefont {M.}~\bibnamefont
  {Rasander}}\ and\ \bibinfo {author} {\bibfnamefont {M.~A.}\ \bibnamefont
  {Moram}},\ }\bibfield  {title} {\bibinfo {title} {On the accuracy of commonly
  used density functional approximations in determining the elastic constants
  of insulators and semiconductors},\ }\href@noop {} {\bibfield  {journal}
  {\bibinfo  {journal} {J. Chem. Phys.}\ }\textbf {\bibinfo {volume} {143}},\
  \bibinfo {pages} {144104} (\bibinfo {year} {2015})}\BibitemShut {NoStop}%
\bibitem [{\citenamefont {Moses}\ \emph {et~al.}(2011)\citenamefont {Moses},
  \citenamefont {Miao}, \citenamefont {Yan},\ and\ \citenamefont {{Van de
  Walle}}}]{MoMi2011}%
  \BibitemOpen
  \bibfield  {author} {\bibinfo {author} {\bibfnamefont {P.~G.}\ \bibnamefont
  {Moses}}, \bibinfo {author} {\bibfnamefont {M.}~\bibnamefont {Miao}},
  \bibinfo {author} {\bibfnamefont {Q.}~\bibnamefont {Yan}},\ and\ \bibinfo
  {author} {\bibfnamefont {C.~G.}\ \bibnamefont {{Van de Walle}}},\ }\bibfield
  {title} {\bibinfo {title} {Hybrid functional investigations of band gaps and
  band alignments for aln, gan, inn, and ingan},\ }\href@noop {} {\bibfield
  {journal} {\bibinfo  {journal} {J. Chem. Phys.}\ }\textbf {\bibinfo {volume}
  {134}},\ \bibinfo {pages} {084703} (\bibinfo {year} {2011})}\BibitemShut
  {NoStop}%
\bibitem [{\citenamefont {Vurgaftman}\ and\ \citenamefont
  {Meyer}(2003)}]{VuMe2003}%
  \BibitemOpen
  \bibfield  {author} {\bibinfo {author} {\bibfnamefont {I.}~\bibnamefont
  {Vurgaftman}}\ and\ \bibinfo {author} {\bibfnamefont {J.}~\bibnamefont
  {Meyer}},\ }\bibfield  {title} {\bibinfo {title} {Band parameters for
  nitrogen-containing semiconductors},\ }\href@noop {} {\bibfield  {journal}
  {\bibinfo  {journal} {J. Appl. Phys.}\ }\textbf {\bibinfo {volume} {94}},\
  \bibinfo {pages} {3675} (\bibinfo {year} {2003})}\BibitemShut {NoStop}%
\bibitem [{\citenamefont {Finn}\ \emph {et~al.}()\citenamefont {Finn},
  \citenamefont {O'Donovan}, \citenamefont {Koprucki},\ and\ \citenamefont
  {Schulz}}]{FiOD2025}%
  \BibitemOpen
  \bibfield  {author} {\bibinfo {author} {\bibfnamefont {R.}~\bibnamefont
  {Finn}}, \bibinfo {author} {\bibfnamefont {M.}~\bibnamefont {O'Donovan}},
  \bibinfo {author} {\bibfnamefont {T.}~\bibnamefont {Koprucki}},\ and\
  \bibinfo {author} {\bibfnamefont {S.}~\bibnamefont {Schulz}},\ }\bibfield
  {title} {\bibinfo {title} {Impact of carrier-density screening on urbach-tail
  energies and optical polarization in (,) quantum well systems},\ }\href
  {https://link.aps.org/doi/10.1103/PhysRevB.103.165201} {\bibfield  {journal}
  {\bibinfo  {journal} {Physical Review Applied}\ }\textbf {\bibinfo {volume}
  {24}},\ \bibinfo {pages} {044084}}\BibitemShut {NoStop}%
\bibitem [{\citenamefont {Dreyer}\ \emph {et~al.}()\citenamefont {Dreyer},
  \citenamefont {Janotti}, \citenamefont {Walle},\ and\ \citenamefont
  {Vanderbilt}}]{DrJo2016}%
  \BibitemOpen
  \bibfield  {author} {\bibinfo {author} {\bibfnamefont {C.~E.}\ \bibnamefont
  {Dreyer}}, \bibinfo {author} {\bibfnamefont {A.}~\bibnamefont {Janotti}},
  \bibinfo {author} {\bibfnamefont {C.~G. V.~d.}\ \bibnamefont {Walle}},\ and\
  \bibinfo {author} {\bibfnamefont {D.}~\bibnamefont {Vanderbilt}},\ }\bibfield
   {title} {\bibinfo {title} {Correct implementation of polarization constants
  in wurtzite materials and impact on {III}-nitrides},\ }\href
  {https://doi.org/10.1103/PHYSREVX.6.021038/SUPPLEMENTALMATERIAL_050416.PDF}
  {\bibfield  {journal} {\bibinfo  {journal} {Physical Review X}\ }\textbf
  {\bibinfo {volume} {6}},\ \bibinfo {pages} {021038}}\BibitemShut {NoStop}%
\bibitem [{\citenamefont {Kobayashi}\ \emph {et~al.}(1983)\citenamefont
  {Kobayashi}, \citenamefont {Sankey}, \citenamefont {Volz},\ and\
  \citenamefont {Dow}}]{KoSa83}%
  \BibitemOpen
  \bibfield  {author} {\bibinfo {author} {\bibfnamefont {A.}~\bibnamefont
  {Kobayashi}}, \bibinfo {author} {\bibfnamefont {O.~F.}\ \bibnamefont
  {Sankey}}, \bibinfo {author} {\bibfnamefont {S.~M.}\ \bibnamefont {Volz}},\
  and\ \bibinfo {author} {\bibfnamefont {J.~D.}\ \bibnamefont {Dow}},\
  }\bibfield  {title} {\bibinfo {title} {Semiempirical tight-binding band
  structures of wurtzite semiconductors: A1n, cds, cdse, zns, and zno},\
  }\href@noop {} {\bibfield  {journal} {\bibinfo  {journal} {Phys. Rev. B}\
  }\textbf {\bibinfo {volume} {28}},\ \bibinfo {pages} {935} (\bibinfo {year}
  {1983})}\BibitemShut {NoStop}%
\bibitem [{\citenamefont {Jogai}(1998)}]{Joga98}%
  \BibitemOpen
  \bibfield  {author} {\bibinfo {author} {\bibfnamefont {B.}~\bibnamefont
  {Jogai}},\ }\bibfield  {title} {\bibinfo {title} {Effect of in-plane biaxial
  strains on the band structure of wurtzite gan},\ }\href@noop {} {\bibfield
  {journal} {\bibinfo  {journal} {Phys. Rev. B}\ }\textbf {\bibinfo {volume}
  {57}},\ \bibinfo {pages} {2382} (\bibinfo {year} {1998})}\BibitemShut
  {NoStop}%
\bibitem [{\citenamefont {O’Reilly}\ \emph {et~al.}(2002)\citenamefont
  {O’Reilly}, \citenamefont {Lindsay}, \citenamefont {Tomic},\ and\
  \citenamefont {Kamal-Saadi}}]{OReLi20021}%
  \BibitemOpen
  \bibfield  {author} {\bibinfo {author} {\bibfnamefont {E.~P.}\ \bibnamefont
  {O’Reilly}}, \bibinfo {author} {\bibfnamefont {A.}~\bibnamefont {Lindsay}},
  \bibinfo {author} {\bibfnamefont {S.}~\bibnamefont {Tomic}},\ and\ \bibinfo
  {author} {\bibfnamefont {M.}~\bibnamefont {Kamal-Saadi}},\ }\bibfield
  {title} {\bibinfo {title} {Tight-binding and k·p models for the electronic
  structure of ga(in)nas and related alloys},\ }\href@noop {} {\bibfield
  {journal} {\bibinfo  {journal} {Semicond. Sci. Technol.}\ }\textbf {\bibinfo
  {volume} {17}},\ \bibinfo {pages} {870} (\bibinfo {year} {2002})}\BibitemShut
  {NoStop}%
\bibitem [{\citenamefont {Boykin}\ \emph {et~al.}(2007)\citenamefont {Boykin},
  \citenamefont {Kharche}, \citenamefont {Klimeck},\ and\ \citenamefont
  {Korkusinski}}]{BoKh2007}%
  \BibitemOpen
  \bibfield  {author} {\bibinfo {author} {\bibfnamefont {T.~B.}\ \bibnamefont
  {Boykin}}, \bibinfo {author} {\bibfnamefont {N.}~\bibnamefont {Kharche}},
  \bibinfo {author} {\bibfnamefont {G.}~\bibnamefont {Klimeck}},\ and\ \bibinfo
  {author} {\bibfnamefont {M.}~\bibnamefont {Korkusinski}},\ }\bibfield
  {title} {\bibinfo {title} {Approximate bandstructures of semiconductor alloys
  from tight-binding supercell calculations},\ }\href@noop {} {\bibfield
  {journal} {\bibinfo  {journal} {J. Phys.: Condens. Matter}\ }\textbf
  {\bibinfo {volume} {19}},\ \bibinfo {pages} {036203} (\bibinfo {year}
  {2007})}\BibitemShut {NoStop}%
\bibitem [{\citenamefont {Bechstedt}\ \emph {et~al.}(2000)\citenamefont
  {Bechstedt}, \citenamefont {Grossner},\ and\ \citenamefont
  {Furthm\"uller}}]{BeGr2000}%
  \BibitemOpen
  \bibfield  {author} {\bibinfo {author} {\bibfnamefont {F.}~\bibnamefont
  {Bechstedt}}, \bibinfo {author} {\bibfnamefont {U.}~\bibnamefont
  {Grossner}},\ and\ \bibinfo {author} {\bibfnamefont {J.}~\bibnamefont
  {Furthm\"uller}},\ }\bibfield  {title} {\bibinfo {title} {Dynamics and
  polarization of group-iii nitride lattices: A first-principles study},\
  }\href {https://doi.org/10.1103/PhysRevB.62.8003} {\bibfield  {journal}
  {\bibinfo  {journal} {Phys. Rev. B}\ }\textbf {\bibinfo {volume} {62}},\
  \bibinfo {pages} {8003} (\bibinfo {year} {2000})}\BibitemShut {NoStop}%
\end{thebibliography}%

\end{document}